\documentclass[aps,pre,showpacs,twocolumn,groupedaddress]{revtex4}

\usepackage{graphicx}

\begin{document}


\title{Analytic Approach for Controlling Quantum States in Complex Systems}


\author{Toshiya Takami}
\affiliation{Computing and Communications Center,
Kyushu University, Fukuoka 812--8581, JAPAN}

\author{Hiroshi Fujisaki}
\affiliation{Department of Chemistry,
Boston University, 590 Commonwealth Ave.,
Boston, Massachusetts, 02215, USA}
\altaffiliation{Present Address:
Institut f\"ur Physikalische und Theoretische Chemie,
J. W. Goethe-Universit\"at, Max-von-Laue-Str. 7,
D-60438 Frankfurt am Main, Germany}


\date{\today}

\begin{abstract}
We examine random matrix systems driven by an external field
in view of optimal control theory (OCT).
By numerically solving OCT equations, 
we can show that there exists a smooth transition 
between two states called ``moving bases'' 
which are dynamically related to 
initial and final states.
In our previous work [J.~Phys.~Soc.~Jpn.~{\bf 73} (2004) 3215-3216; 
Adv.~Chem.~Phys.~{\bf 130A} (2005) 435-458],
they were assumed to be orthogonal,
but in this paper, we introduce orthogonal moving bases. 
We can construct a Rabi-oscillation like representation 
of a wavpacket using such moving bases, and derive an analytic 
optimal field as a solution of the OCT equations. 
We also numerically show that the newly obtained optimal field 
outperforms the previous one.
\end{abstract}

\pacs{05.45.Mt, 02.30.Yy, 03.65.Sq}

\maketitle


\section{Introduction\label{sec:intro}}

Controlling atomic and molecular processes by laser fields is
one of current topics in physics and chemistry \cite{RZ00}.
There are various control schemes applied to such processes:
$\pi$-pulses \cite{AE87}, nonadiabatic transitions \cite{TN98}, 
adiabatic rapid passage \cite{MGHGW94}, 
Stimulated Raman Adiabatic Passage (STIRAP) \cite{BTS98,KAS02,GR04},
pulse-timing control \cite{TR85}, 
and coherent control \cite{SB03}, etc.
These strategies are known to work when the 
system to be controlled is rather simple or small.
However, the system can be complex \cite{Gutzwiller90} when we deal with
highly excited states in large molecules or mesoscopic devices 
driven by electro-magnetic fields.
Such a ``complex'' system in the limit of strong chaos is modelled
by a random matrix Hamiltonian with a time-dependent external field
\cite{Haake01},
and the dynamics is well
represented by multi-level-multi-level transitions with  
random interactions among energy levels.
Although there are many works on the statistical properties
\cite{GRMN90,TKM91,TH92,ZD93,TMO07}
as well as the semiclassical properties \cite{TKM92,TKM94,TKM95}
of eigenvalues under the variation of an external parameter,
few works have been published on the dynamical properties of such systems
except several studies on nonadiabatic processes \cite{Wilkinson88,WM00}.

Even for such complex systems, there exist mathematical results
showing complete controllability \cite{HTC83,RHR04} of general quantum systems
with discrete spectrum under certain conditions.
The existence of an optimal field is proved
by optimal control theory (OCT) \cite{PDR88},
which is a powerful tool to obtain an optimal field
and has been studied for various dynamical systems \cite{AOFD05,SMY05}.
For the purpose of steering quantum states,
many numerical schemes with monotonically convergent algorithms
\cite{RZ00,ZBR98,OZR99} have been developed based on OCT.
In general, OCT for quantum states provides sets of nonlinear
differential equations (OCT equations) which are solved by iterative procedures.
For complex systems with many degrees of freedom, however,
the optimal field often becomes too complicated
to analyze the dynamical processes involved.
In addition, the computational cost becomes significantly heavy
when we apply OCT to realistic problems with many degrees of freedom.
Analytic approaches can be a good strategy to complement this annoying 
situation.

One such analytic method for multi-level control problems
is STIRAP \cite{BTS98,KAS02,GR04}.
Though it can accomplish perfect control,
it assumes an intermediate state coupled to initial and target states,
and uses a pair of external fields with slowly varying amplitudes.
Recently, we have proposed another analytic optimal field \cite{TF04,TFM05}
which induces a ``direct'' transition between 
random vectors in a random matrix system. 
The key idea of this approach is to describe
the optimally controlled dynamics as a Rabi-like oscillation \cite{AE87},
and our optimal field can be interpreted as a generalized $\pi$-pulse
\cite{TF04}.
Though the derivation and applicability of our analytic optimal field
have been detailed in \cite{TFM05},
there exists deficiency in our previous formula because of several
(unnecessary) assumptions for simplification. 
In this paper, we rederive an analytic optimal field with less numbers of 
assumptions and reexamine its applicability to random matrix systems.

This paper is organized as follows.
In Sec.~\ref{sec:numerical},
we numerically investigate the multi-state control problem
by OCT to show that, in some cases,
the optimal field induces a smooth transition.
According to this observation,
in section \ref{sec:analytic},
we introduce a Rabi-like representation of the controlled state
with some modifications compared to our previous result \cite{TF04,TFM05}.
Employing this representation,
we obtain a new analytic expression of the optimal field.
In Sec.~\ref{sec:application},
we confirm the applicability of the analytic field
through the numerical integration of the Schr\"odinger's equation 
for random matrix systems.
Finally, in Sec.~\ref{sec:summary}, 
we summarize this paper and give some discussions 
on the control problem of quantum chaos systems.
We mention some technical details in Appendix.

\section{Optimal Control in Random Matrix Systems\label{sec:numerical}}

We present numerical results
of controlled dynamics driven by an optimal field
to see what kinds of dynamics are involved in random matrix systems.
The random matrix Hamiltonian driven by a time-dependent
external field $\varepsilon(t)$ is written as
\begin{equation}
\label{eqn:RM-Hamiltonian}
  H[\varepsilon(t)]=H_0+\varepsilon(t)V,
\end{equation}
where $H_0$ and $V$ are random matrices subject to a certain universality
class \cite{Haake01}, i.e. Gaussian Orthogonal Ensemble (GOE),
Gaussian Unitary Ensemble (GUE), etc.

It is well known that a strongly chaotic system does not have
any constant of motion except the total energy \cite{Gutzwiller90},
where the typical quantum states are random vectors.
Thus, it is appropriate to choose initial and target states as random vectors.
If we choose a certain ortho-normalized basis,
a random vector in $N$-dimensional Hilbert space is
represented by a set of random complex numbers $\{c_j\}$.
If such a vector has neither special symmetry nor correlation,
only the constraint imposed is the normalization condition,
\begin{equation}
  \sum_{j=1}^N|c_j|^2=1.
\end{equation}
Then, the normalized probability density for a variable $y=|c_j|^2$ is given by
\begin{equation}
\label{eqn:random-vector}
  P_N(y)\ dy=N\exp(-Ny)\ dy,
\end{equation}
when $N$ is sufficiently large \cite{Haake01}.

The actual procedure to numerically obtain an optimal field is as follows:
The Hamiltonian (\ref{eqn:RM-Hamiltonian}) is constructed
by generating two random matrices, $H_0$ and $V$, with $N\times N$ elements,
where the scales of them are determined so that
the averaged eigenvalue-spacing $\Delta E$ of $H_0$ and the variance
$\Delta V$ of off-diagonal elements of $V$ are both unity.
Next, we define an initial state $|\Phi_0\rangle$ and
a target state $|\Phi_T\rangle$
as random vectors satisfying the distribution (\ref{eqn:random-vector}).
Then, for a fixed target time $T$,
the optimal field $\varepsilon(t)$ is obtained 
by solving the OCT equations which are detailed in Sec.~\ref{sec:ZBR-OCT}.

\subsection{Zhu-Botina-Rabits Scheme of OCT\label{sec:ZBR-OCT}}

There are many effective methods to solve OCT equations 
for quantum systems \cite{RZ00}. 
In this section,
we use a method introduced by Zhu, Botina, and Rabitz \cite{ZBR98}
(ZBR-OCT).
Our goal is to determine the optimal external
field $\varepsilon(t)$ by which a given initial state $|\Phi_0\rangle$ is
steered to a given target state $|\Phi_T\rangle$ at a target time $T$.
According to ZBR-OCT,
we introduce a functional $J(\varepsilon(t),|\phi(t)\rangle)$
\begin{widetext}
\begin{equation}
  J(\varepsilon(t),|\phi(t)\rangle)
  =J_0-\alpha\int_0^T\left[\varepsilon(t)\right]^2dt
  -2{\rm Re}\left[\langle\phi(T)|\Phi_T\rangle
    \int_0^T\langle\chi(t)|{\partial\over\partial t}
       -{H[\varepsilon(t)]\over i\hbar}|\phi(t)\rangle dt\right],
\label{eqn:J}
\end{equation}
\end{widetext}
where $T$ and $\alpha$ are given parameters representing
the target time and the penalty factor, respectively.
The quantum state $|\phi(t)\rangle$ satisfies the initial condition,
$|\phi(0)\rangle=|\Phi_0\rangle$.
The first term in the right-hand side is the final overlap,
\begin{equation}
\label{eqn:J0}
  J_0=\left|\langle\phi(T)|\Phi_T\rangle\right|^2.
\end{equation}
The second term is the penalty term which minimizes the amplitude
of the optimal field.
In the third term, a Lagrange multiplier $|\chi(t)\rangle$ is
introduced to give a constraint
that $|\phi(t)\rangle$ satisfies Schr\"odinger's equation,
\begin{equation}
\label{eqn:Schroedinger-phi}
  i\hbar{d\over dt}|\phi(t)\rangle
  =H[\varepsilon(t)]|\phi(t)\rangle.
\end{equation}
On the other hand, Schr\"odinger's equation for $|\chi(t)\rangle$ is
\begin{equation}
\label{eqn:Schroedinger-chi}
  i\hbar{d\over dt}|\chi(t)\rangle
  =H[\varepsilon(t)]|\chi(t)\rangle,
\end{equation}
and the boundary condition $|\chi(T)\rangle=|\Phi_T\rangle$
are obtained by ``differentiating'' the functional with 
respect to $|\phi(t)\rangle$ and $|\phi(T)\rangle$.
For the Hamiltonian (\ref{eqn:RM-Hamiltonian}),
the variation of $J$ with respect to $\varepsilon(t)$
gives an expression for the optimal field,
\begin{equation}
\label{eqn:ZBR-field}
  \varepsilon(t)
  ={1\over\alpha\hbar}{\rm Im}\left[\langle\phi(t)|\chi(t)\rangle
     \langle\chi(t)|V|\phi(t)\rangle\right].
\end{equation}
This is a self-consistent expression for the optimal field, 
and to obtain its actual value, 
we have to simultaneously solve the nonlinear coupled equations,
(\ref{eqn:Schroedinger-phi}),
(\ref{eqn:Schroedinger-chi}) and (\ref{eqn:ZBR-field}) (OCT equations). 
ZBR-OCT is one such method which numerically solves the OCT 
equations with iterative procedures \cite{ZBR98}.

\begin{figure*}
  \includegraphics[scale=0.5]{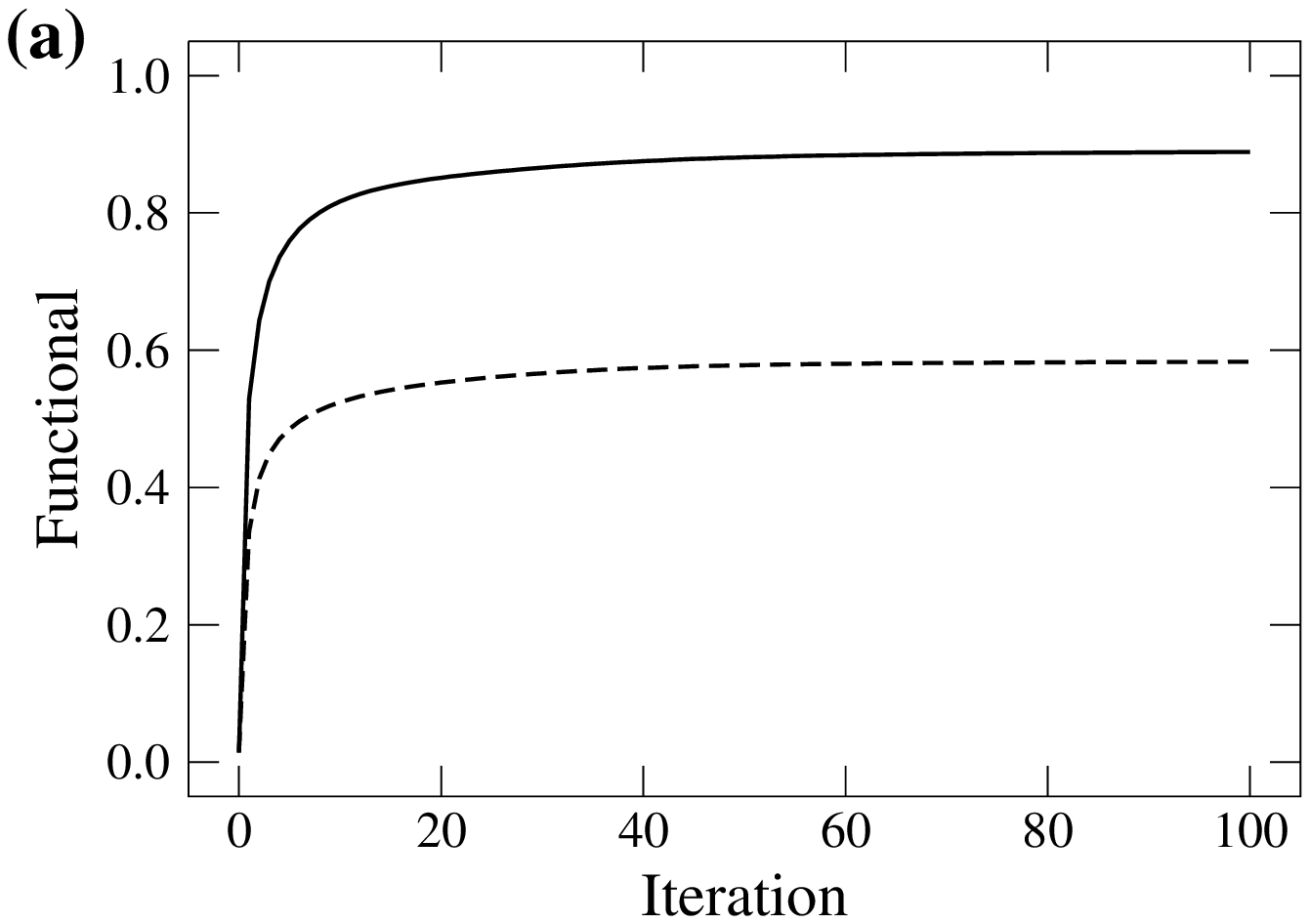}%
  \includegraphics[scale=0.5]{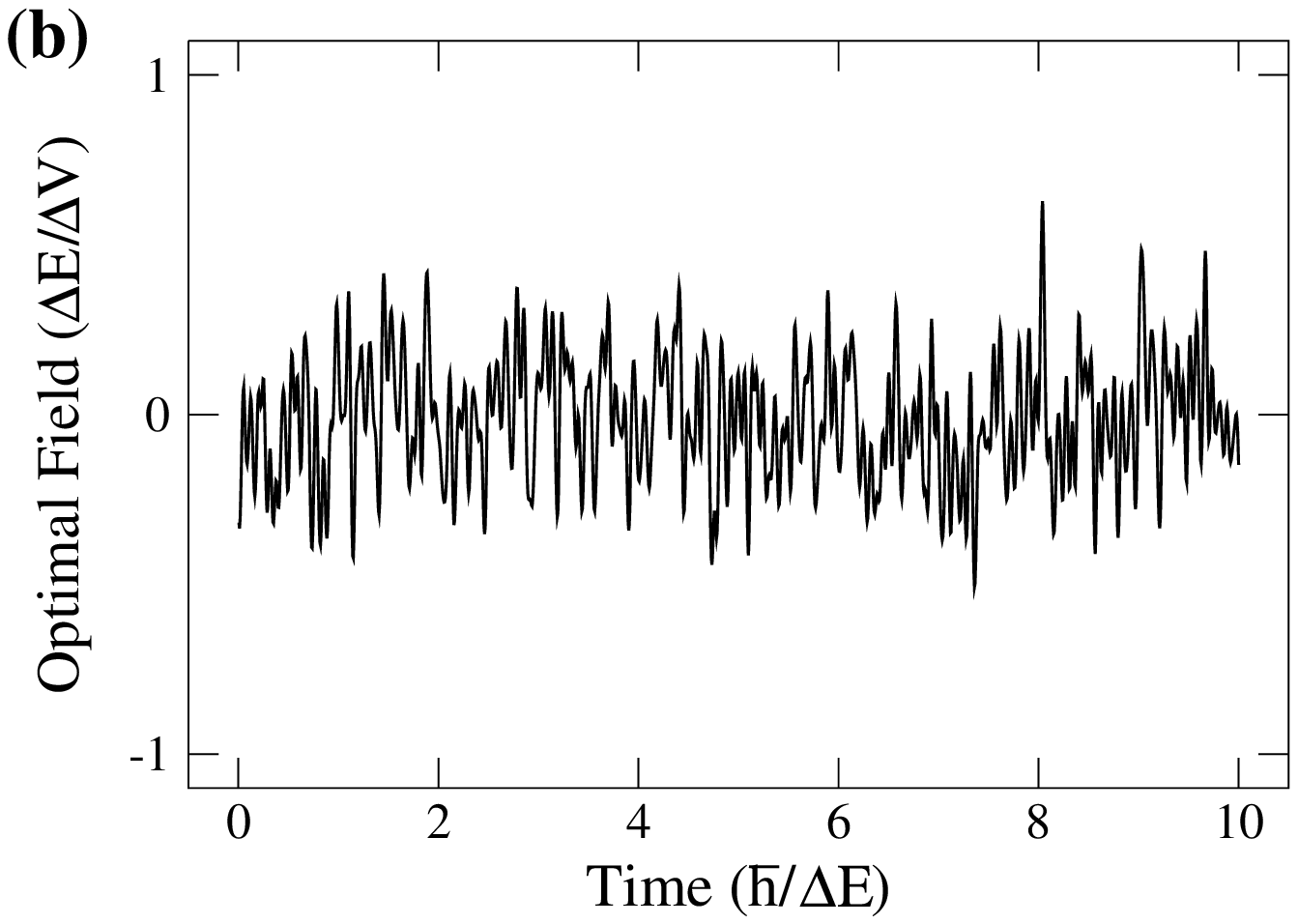}%
\\
  \includegraphics[scale=0.5]{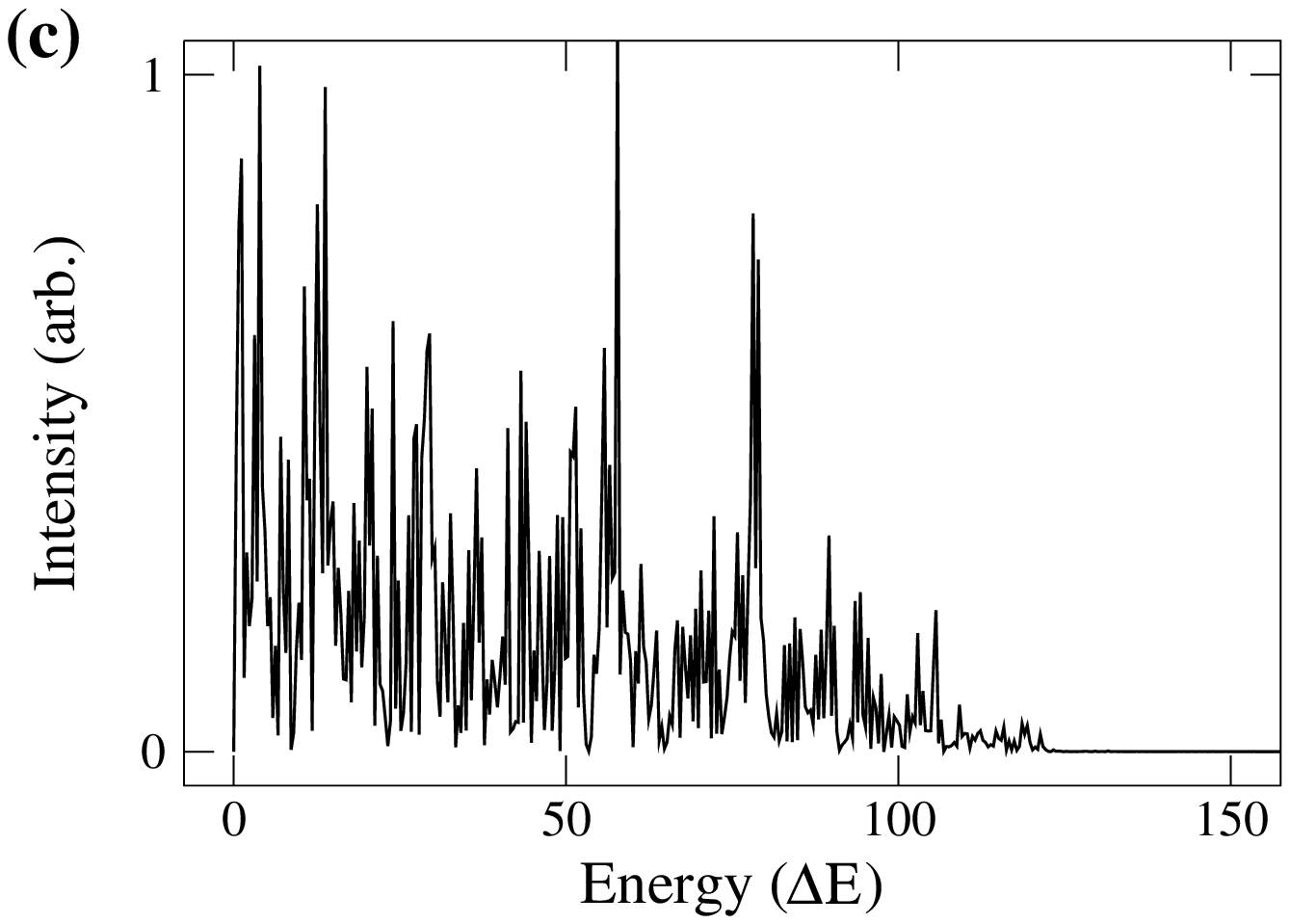}%
  \includegraphics[scale=0.5]{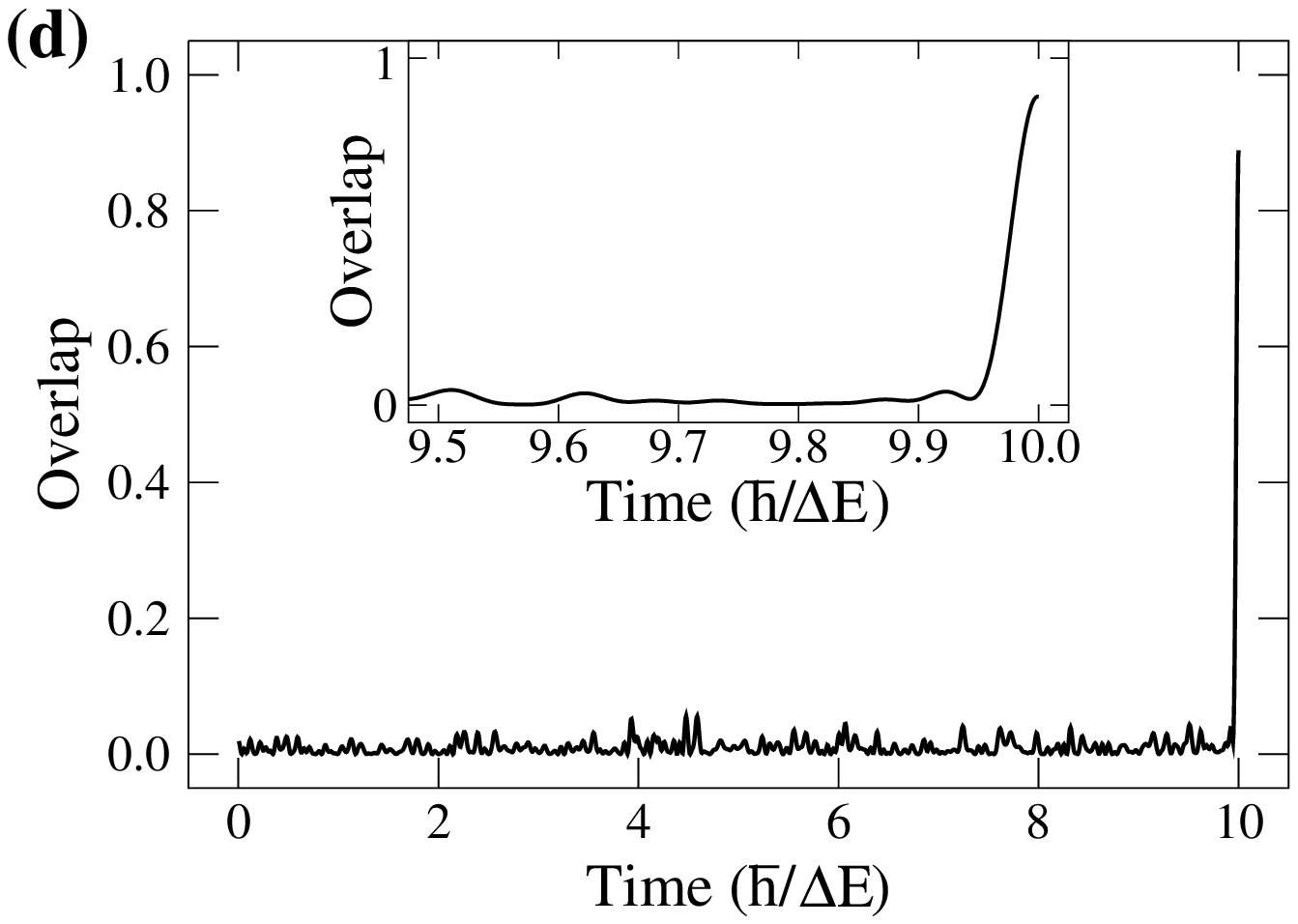}%
\caption{\label{fig:numerical-short}
Numerical results of controlled dynamics between random vectors
in a $128\times128$ GOE random matrix system.
The optimal field is obtained through the iterative procedure
given by Zhu, Botina, and Rabitz \cite{ZBR98}
for the target time $T=10$ and the penalty factor $\alpha=1$.
(a) Convergence property of functional values,
$J_0$ (solid curve) and $J$ (dashed curve),
as a function of the iteration step.
(b) Optimal external field $\varepsilon(t)$.
(c) Power spectrum of $\varepsilon(t)$.
(d) Time evolution of the overlap $|\langle\Phi_T|\psi(t)\rangle|^2$
(the magnified curve near $t=T$ is shown in the inset).
}
\end{figure*}

\begin{figure*}
  \includegraphics[scale=0.5]{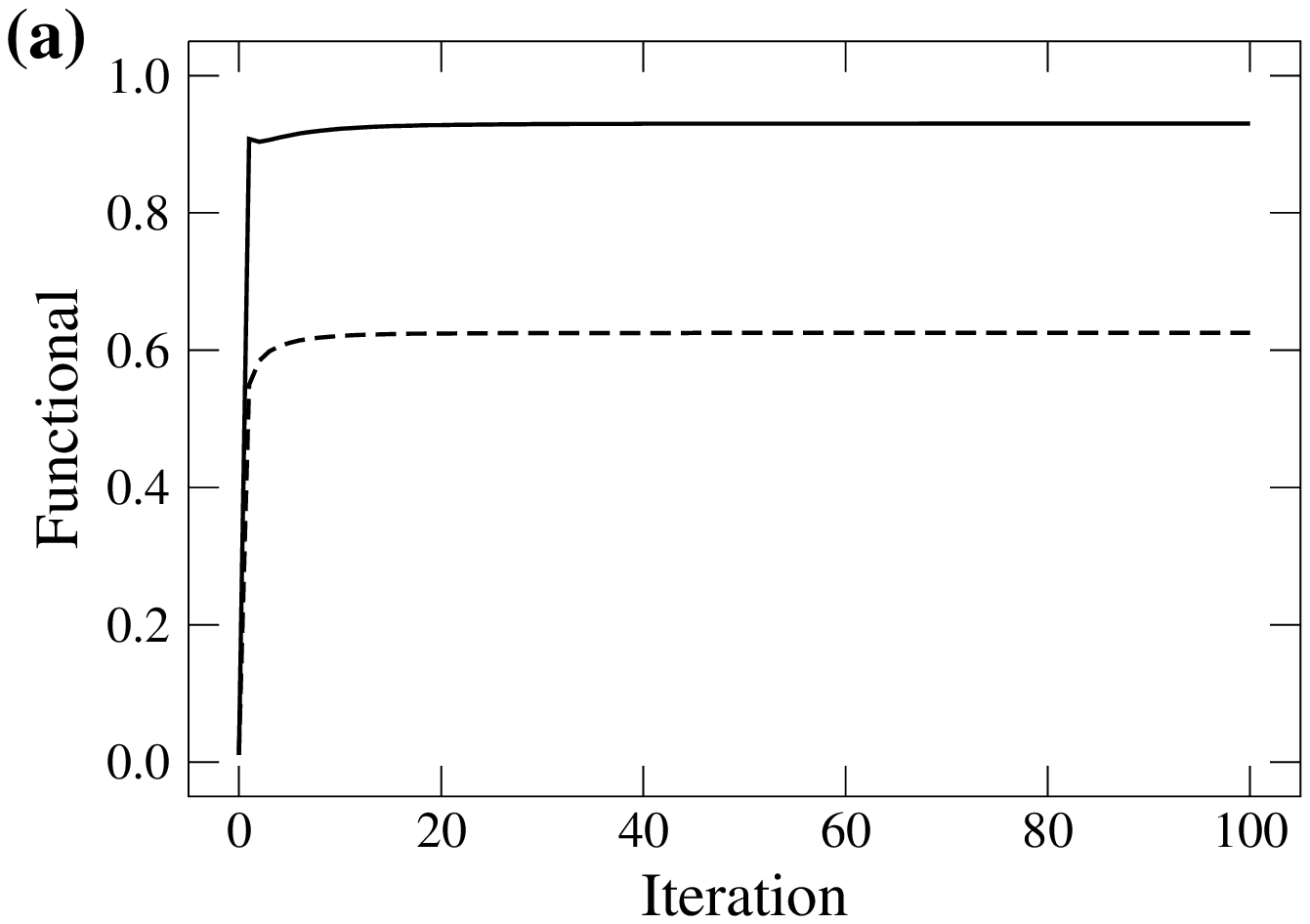}%
  \includegraphics[scale=0.5]{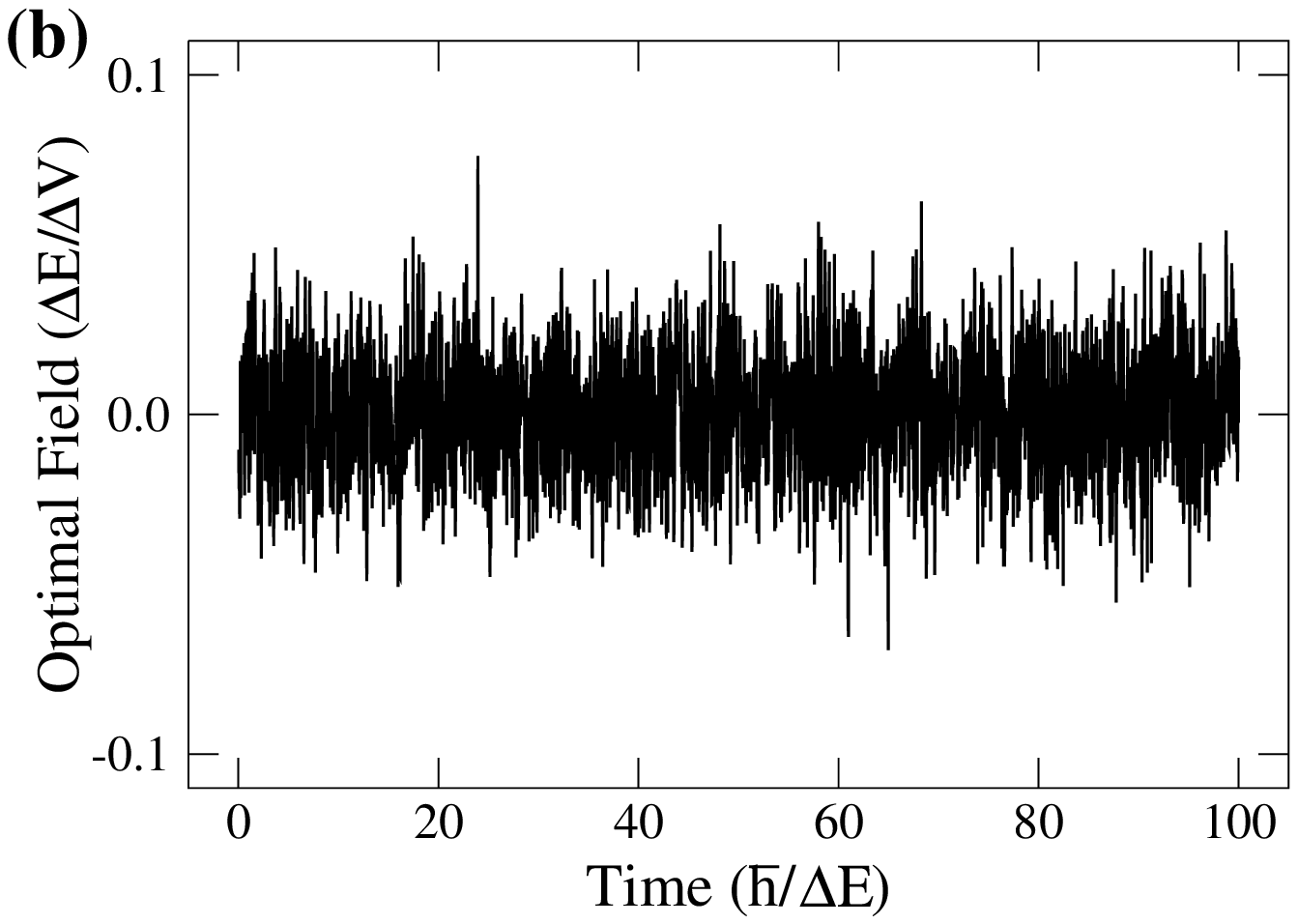}%
\\
  \includegraphics[scale=0.5]{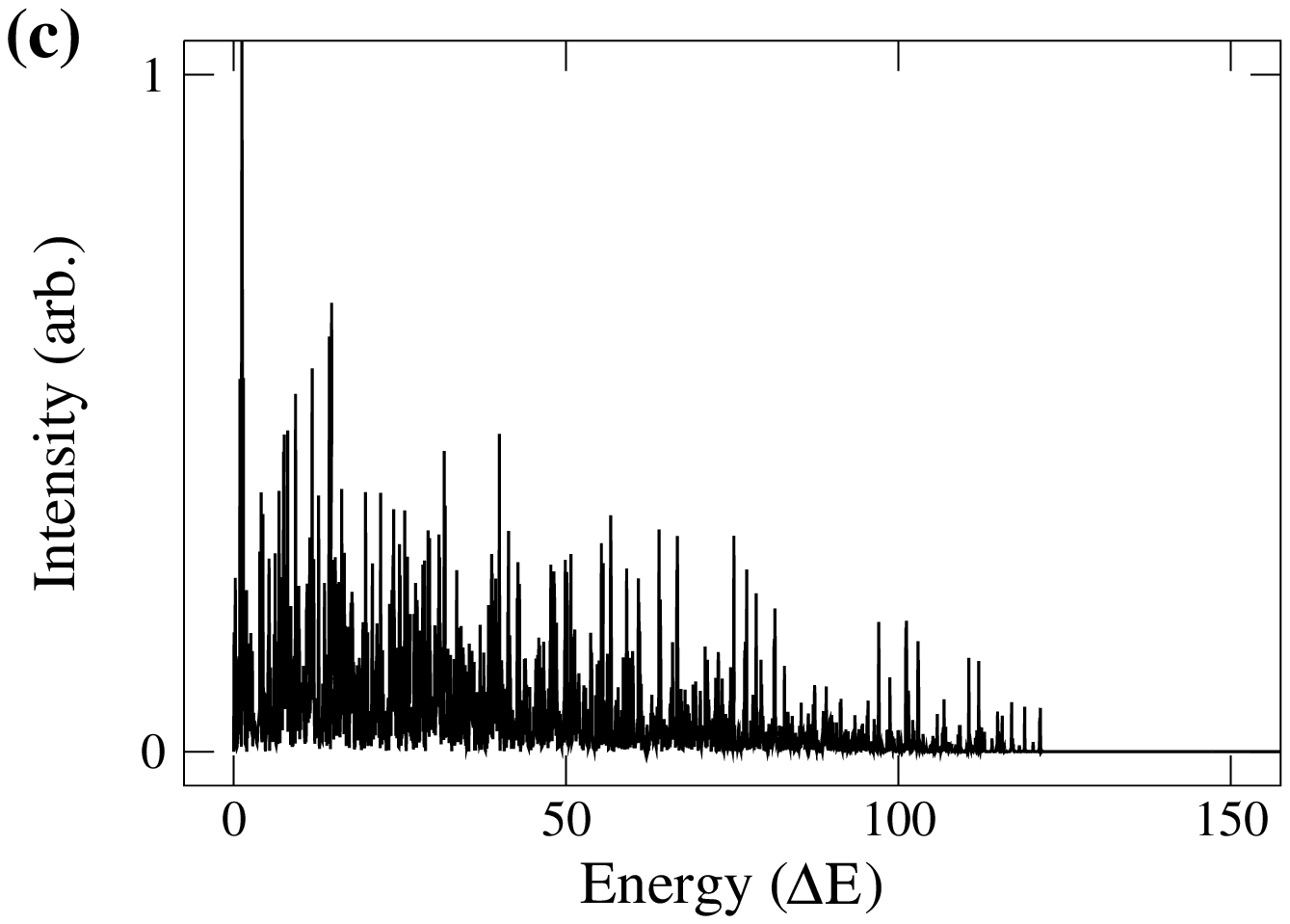}%
  \includegraphics[scale=0.5]{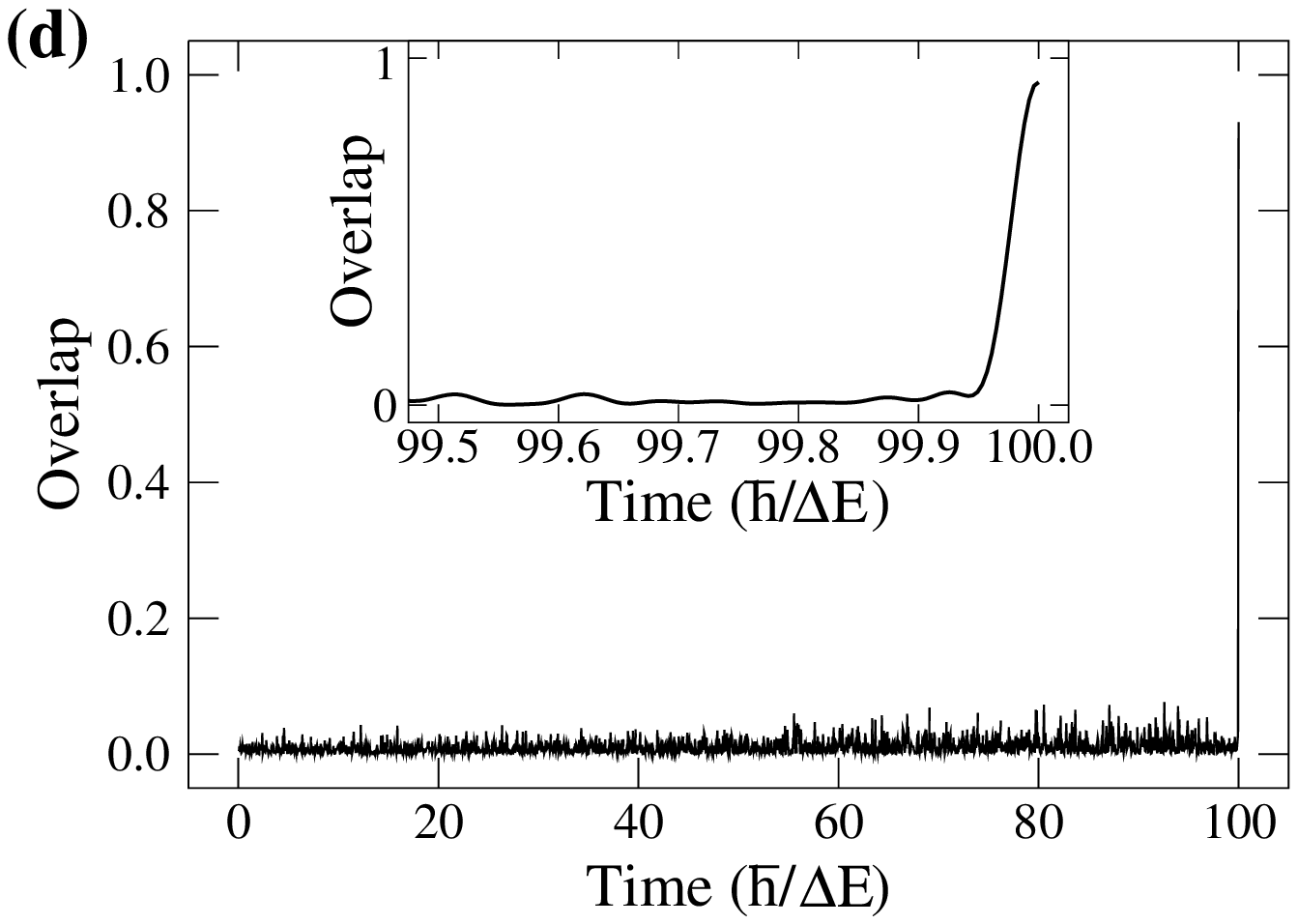}%
\caption{\label{fig:numerical-long}
The same as Fig.~\ref{fig:numerical-short} except 
that the target time is longer ($T=100$) and the penalty 
factor is larger ($\alpha=10$).
}
\end{figure*}

\begin{figure*}
  \includegraphics[scale=0.5]{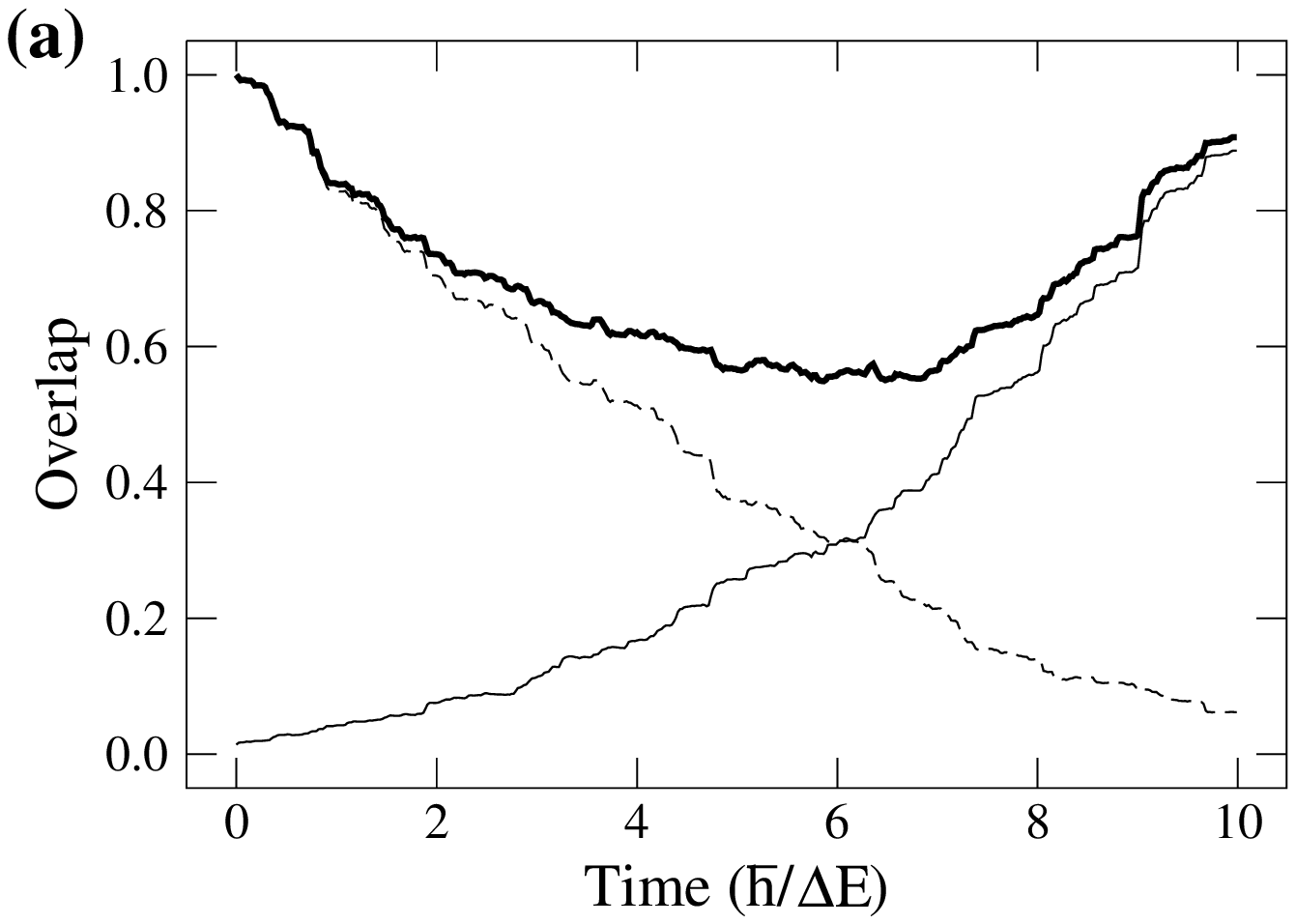}%
  \includegraphics[scale=0.5]{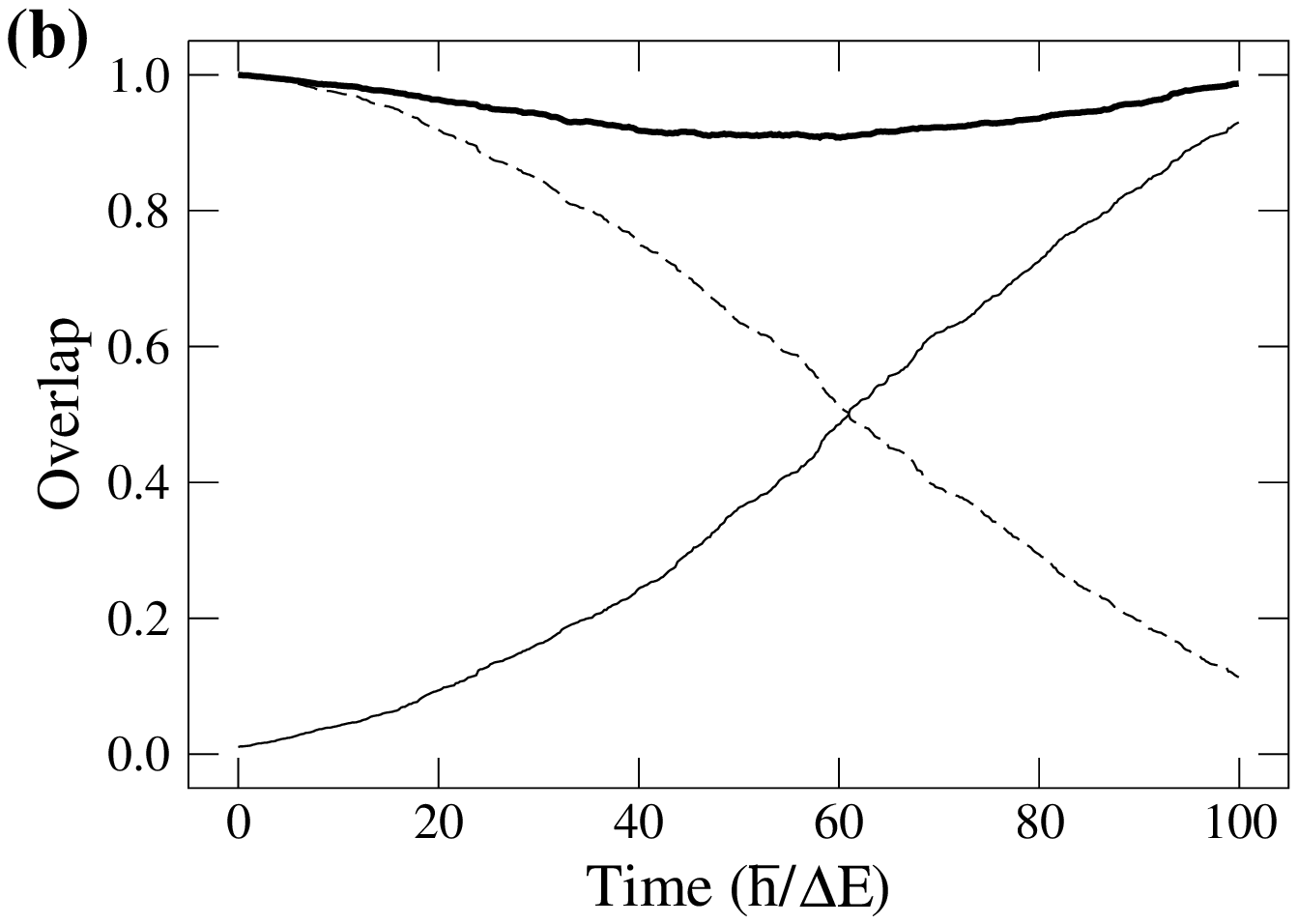}%
\caption{\label{fig:numerical-smooth}
Time evolution of the overlaps
$\langle\hat P_\chi(t)\rangle$ (solid curve),
$\langle\hat P_\phi(t)\rangle$ (dashed curve), and
$\langle\hat P_{\phi+\chi}(t)\rangle$ (thick curve) defined by 
Eqs.~(\ref{eq:proj1}), (\ref{eq:proj2}) and (\ref{eq:proj3}).
(a) and (b) correspond to Fig.~\ref{fig:numerical-short} ($T=10$ and $\alpha=1$)
and Fig.~\ref{fig:numerical-long} ($T=100$ and $\alpha=10$), respectively.
}
\end{figure*}

\subsection{Numerical Results\label{sec:oct-numerical}}

We show controlled dynamics driven by
numerically obtained optimal fields 
for a $128\times128$ GOE random matrix Hamiltonian.
The quantum state $|\psi(t)\rangle$ with the initial condition
$|\psi(0)\rangle=|\Phi_0\rangle$ evolves
according to Schr\"odinger's equation (\ref{eqn:Schroedinger-phi}).
We have chosen $H_0$ and $V$ so that $\Delta E=\Delta V=1$.
In other words, the energy values are shown in unit of $\Delta E$,
and the unit of time is $\hbar/\Delta E$.
Then, the field strength $\varepsilon(t)$ is
shown in unit of $\Delta E/\Delta V$.

In Fig.~\ref{fig:numerical-short},
we show the result with the parameters $T=10$ and $\alpha=1$.
The target time $T=10$ is comparable to
the minimum time $\tau_0\approx2\pi$ ($\hbar/\Delta E$)
which is necessary to resolve each energy level from its adjacent levels.
In Fig.~\ref{fig:numerical-short}(a),
the functional values $J_0$ (\ref{eqn:J0}) and
$J$ (\ref{eqn:J}) are shown as the solid and dashed curves, respectively.
They appear to converge after several ten steps.
The final overlap $J_0$ is $0.89$ after $100$ iterations.
The optimal field $\varepsilon(t)$
is shown in Fig.~\ref{fig:numerical-short}(b)
as well as its Fourier spectrum in Fig.~\ref{fig:numerical-short}(c).
Figure~\ref{fig:numerical-short}(d) shows
the time evolution of the overlap $|\langle\Phi_T|\psi(t)\rangle|^2$
with its magnification near the target time in the inset.

Figure~\ref{fig:numerical-long} shows
the result obtained for the parameters $T=100$ and $\alpha=10$,
which is the case of a relatively long target time compared to $\tau_0$.
The values of $J_0$ and $J$, the optimal field $\varepsilon(t)$,
its Fourier spectrum,
and the overlap $|\langle\Phi_T|\psi(t)\rangle|^2$ are shown as in Fig.~\ref{fig:numerical-short}.
In this calculation, the final overlap $J_0$ is $0.93$ after $100$ iterations.

In both cases, the overlap $|\langle\Phi_T|\psi(t)\rangle|^2$
as a function of time $t$ remains small
until $t$ is close to the target time $T$.
In multi-state quantum dynamics, 
even if no external field is applied,
an auto-correlation function $\langle\psi(0)|\psi(t)\rangle$ 
can rapidly decay by dephasing among dynamical phases of $H_0$.
This is the reason why the overlaps $\langle \Phi_T| \psi(t) \rangle$
in Fig.~\ref{fig:numerical-short}(d) and Fig.~\ref{fig:numerical-long}(d)
rapidly grow up to the final values near $t=T$.
In other words, $\langle \Phi_T| \psi(t) \rangle$ decays quickly
when $t$ deviates from $T$.

\subsection{Observation of Smooth Transitions\label{sec:observe}}

Since we want to concentrate on transitions induced by $\varepsilon(t)$ only,
it is necessary to remove the contribution from dephasing by $H_0$.
This is nothing but the procedure of the interaction picture in 
quantum mechanics \cite{J.J.Sakurai94}.
We define time-dependent quantum states
related to $|\Phi_0\rangle$ and $|\Phi_T\rangle$ by
\begin{equation}
\label{eqn:time-dependent-states}
  |\phi_0(t)\rangle=\hat U_0(t,0)|\Phi_0\rangle,\quad
  |\chi_0(t)\rangle=\hat U_0(t,T)|\Phi_T\rangle,
\end{equation}
where $\hat U_0(t_2,t_1)$ represents a propagator
from $t=t_1$ to $t=t_2$ with respect to the unperturbed Hamiltonian $H_0$.
We call these states ``moving bases.''
In the following, we analyze the optimally controlled dynamics
through these time-dependent states.

If we introduce projection operators associated with these states by
\begin{equation}
  \hat P_\phi(t)=|\phi_0(t)\rangle\langle\phi_0(t)|,\quad
  \hat P_\chi(t)=|\chi_0(t)\rangle\langle\chi_0(t)|,
\end{equation}
the probabilities such that $|\psi(t) \rangle$ is found in these states 
are written as
\begin{eqnarray}
\label{eq:proj1}
  \langle\hat P_\phi(t)\rangle
    \equiv\langle\psi(t)|\hat P_\phi|\psi(t)\rangle,\\
\label{eq:proj2}
  \langle\hat P_\chi(t)\rangle
    \equiv\langle\psi(t)|\hat P_\chi|\psi(t)\rangle.
\end{eqnarray}
These values are more appropriate quantities
to observe the multi-level-multi-level transition dynamics
compared to the bare overlap $|\langle\Phi_T|\psi(t)\rangle|^2$ as shown below.
In addition, we introduce another projection operator
\begin{equation}
  \hat P_{\phi+\chi}(t)
  ={\hat P_\phi(t)+\hat P_\chi(t)
     -\hat P_\phi(t)\hat P_\chi(t)-\hat P_\chi(t)\hat P_\phi(t)
    \over1-{\rm tr}[\hat P_\chi(t)\hat P_\phi(t)]},
\end{equation}
which represents projection onto a subspace defined by a linear
superposition of $|\phi_0(t)\rangle$ and $|\chi_0(t)\rangle$.
We can prove that this is a projection operator
by using $1-{\rm tr}[\hat P_\phi(t) \hat P_\chi(t)]
=1-|\langle \phi_0|\chi_0 \rangle|^2$
and $\hat P_\phi(t) \hat P_\chi(t) \hat P_\phi(t)
=|\langle \phi_0|\chi_0 \rangle|^2 \hat P_\phi(t)$, etc.
Then, the quantity,
\begin{equation}
  \langle\hat P_{\phi+\chi}(t)\rangle
  \equiv\langle\psi(t)|\hat P_{\phi+\chi}(t)|\psi(t)\rangle,
\label{eq:proj3}
\end{equation}
represents the probability that the quantum state $|\psi(t)\rangle$ 
is found on the subspace.

In Fig.~\ref{fig:numerical-smooth}(a) and (b),
we show the overlaps (probabilities),
$\langle\hat P_\phi(t)\rangle$, $\langle\hat P_\chi(t)\rangle$,
and $\langle\hat P_{\phi+\chi}(t)\rangle$, calculated from the results 
in Figs.~\ref{fig:numerical-short} and \ref{fig:numerical-long}.
All the curves in Fig.~\ref{fig:numerical-smooth}(b) are smoother 
than those in Fig.~\ref{fig:numerical-smooth}(a).
It is also worth noting that $\langle\hat P_{\phi+\chi}(t)\rangle$ stays
close to unity for all the time in Fig.~\ref{fig:numerical-smooth}(b).

From other ZBR-OCT calculations for random matrix systems,
we found that the ZBR optimal field 
induces a transition from $|\Phi_0\rangle$ to $|\Phi_T\rangle$
nearly within a subspace spanned by $|\phi_0(t)\rangle$ and $|\chi_0(t)\rangle$
when the target time $T$ is sufficiently large.
Based on this finding,
we will develop an analytic approach for the optimal field
in the next section.

\section{Analytic Approach for Controlled Dynamics\label{sec:analytic}}

The Rabi oscillation in a two-level system has been studied in detail \cite{AE87}.
According to such previous works, we can represent
a wavefunction as a linear combination of two eigenstates
$|\varphi_1\rangle$ and $|\varphi_2\rangle$
with eigen-energies $E_1$ and $E_2$,
\begin{equation}
  |\psi(t)\rangle=A(t)|\varphi_1\rangle e^{E_1t/i\hbar}
                 +B(t)|\varphi_2\rangle e^{E_2t/i\hbar}.
\end{equation}
Here the coefficients $A(t)$ and $B(t)$ are slowly oscillating functions
with a Rabi frequency under the rotating-wave approximation (RWA) \cite{AE87}.

In this section, 
we show, under certain conditions, that a Rabi-like description
becomes valid even for multi-level quantum systems
where the wavefunction is described by the time-dependent states in Eq.~(\ref{eqn:time-dependent-states}). 
We call these states ``moving bases'' instead of eigenstates.
This is equivalent to considering the case where the controlled state remains
in the subspace spaned by these moving bases
over a whole period of the dynamics.
With the help of OCT,
we conversely obtain an analytical expression for
the optimal field to induce the smooth transition 
we found in Fig.~\ref{fig:numerical-smooth} for the multi-level dynamics.

\subsection{Rabi-like Representation}

In the previous section,
we have defined the moving bases
$|\phi_0(t)\rangle$ and $|\chi_0(t)\rangle$ 
(\ref{eqn:time-dependent-states})
in order to observe smooth transitions in OCT calculations.
Unlike the two-level case, however, these states
are not always orthogonal to each other.
The inner-product between them is written in the form,
\begin{equation}
\label{eqn:inner-product}
  \langle\phi_0(t)|\chi_0(t)\rangle
  =\langle\Phi_0|\hat U(0,T)|\Phi_T\rangle
  =ie^{i\theta}\sin\Theta,
\end{equation}
with $0\le\Theta\le\pi/2$ and $0\le\theta<2\pi$.
Note that
the inner-product (\ref{eqn:inner-product}) does not depend on $t$,
i.e. constant in time, despite that
the moving bases ($|\phi_0(t)\rangle$ and $|\chi_0(t)\rangle$) 
rapidly change their ``directions'' according to the Schr\"odinger's equations.

In our previous works \cite{TF04,TFM05}, 
we have used an assumption that $\Theta =0$.
Actually we can remove this assumption by introducing
an orthogonal pair of the moving bases as (Schmidt decomposition)
\begin{eqnarray}
\label{eqn:tilde_phi}
  |\tilde\phi_0(t)\rangle&=&|\phi_0(t)\rangle,\\
\label{eqn:tilde_chi}
  |\tilde\chi_0(t)\rangle&=&{|\chi_0(t)\rangle
    -ie^{i\theta}\sin\Theta|\phi_0(t)\rangle\over\cos\Theta}.
\end{eqnarray}
These are our new moving bases which are orthogonal to each other 
and will be used below.

We introduce a Rabi-like representation of the quantum state $|\phi(t)\rangle$
with an initial condition $|\phi(0)\rangle=|\Phi_0\rangle$ by a linear
combination of the new moving bases $|\tilde\phi_0(t)\rangle$ and $|\tilde\chi_0(t)\rangle$,
\begin{equation}
\label{eqn:two-state}
  |\phi(t)\rangle
  =A(t)|\tilde\phi_0(t)\rangle+B(t)|\tilde\chi_0(t)\rangle.
\end{equation}
The coefficients, $A(t)$ and $B(t)$, 
must satisfy a normalization condition,
\begin{equation}
  |A(t)|^2+|B(t)|^2=1.
\end{equation}
If this representation is valid,
$A(t)$ and $B(t)$ satisfy the following differential equations,
\begin{widetext}
\begin{eqnarray}
\label{eqn:de-A}
  {d\over dt}A(t)
  &=&{\varepsilon(t)\over i\hbar}\left[
     \langle\tilde\phi_0(t)|V|\tilde\phi_0(t)\rangle A(t)
    +\langle\tilde\phi_0(t)|V|\tilde\chi_0(t)\rangle B(t)
  \right],\\
\label{eqn:de-B}
  {d\over dt}B(t)
  &=&{\varepsilon(t)\over i\hbar}\left[
     \langle\tilde\chi_0(t)|V|\tilde\phi_0(t)\rangle A(t)
    +\langle\tilde\chi_0(t)|V|\tilde\chi_0(t)\rangle B(t)
  \right].
\end{eqnarray}
\end{widetext}

\begin{figure*}
 \includegraphics[scale=0.5]{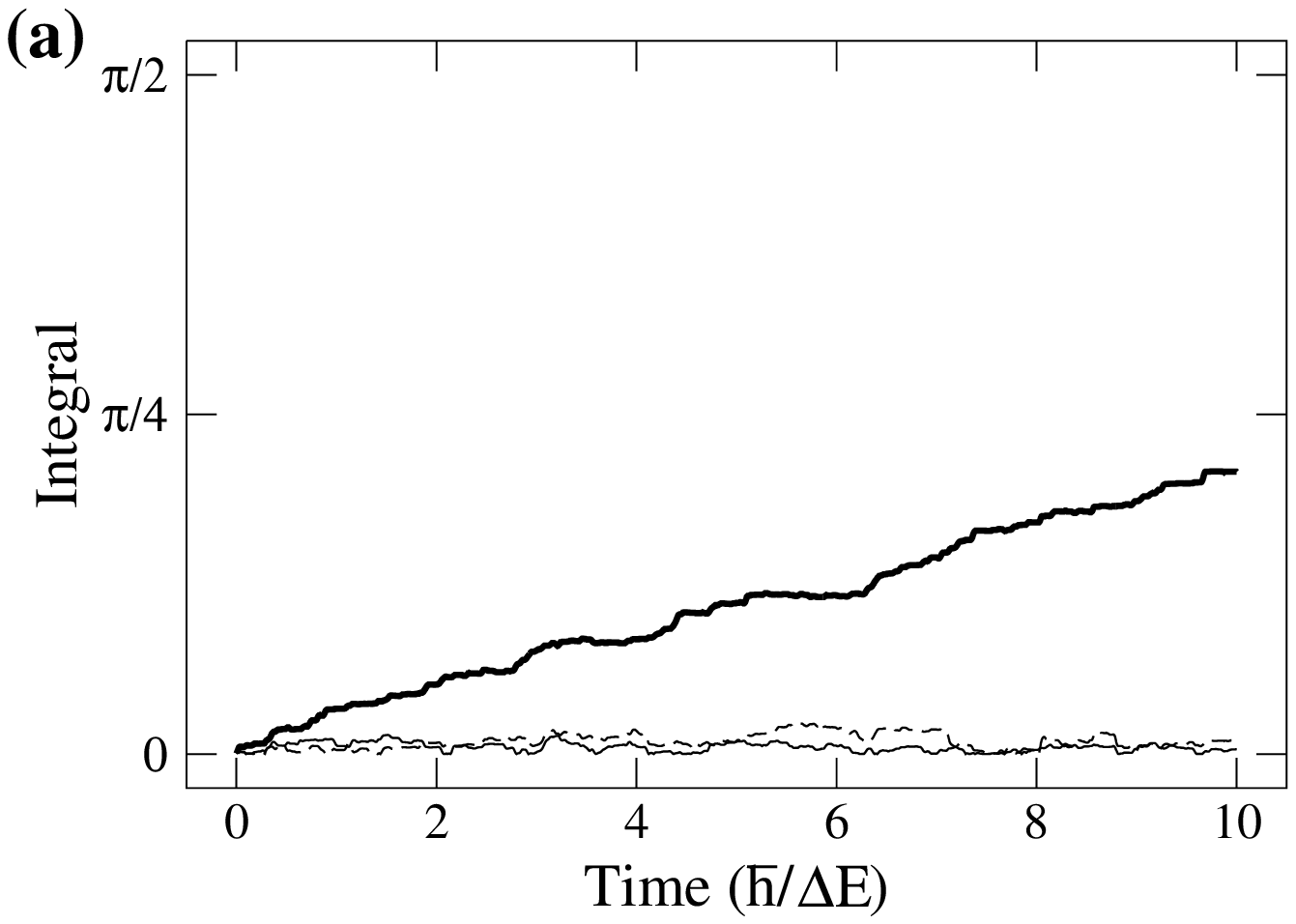}
 \includegraphics[scale=0.5]{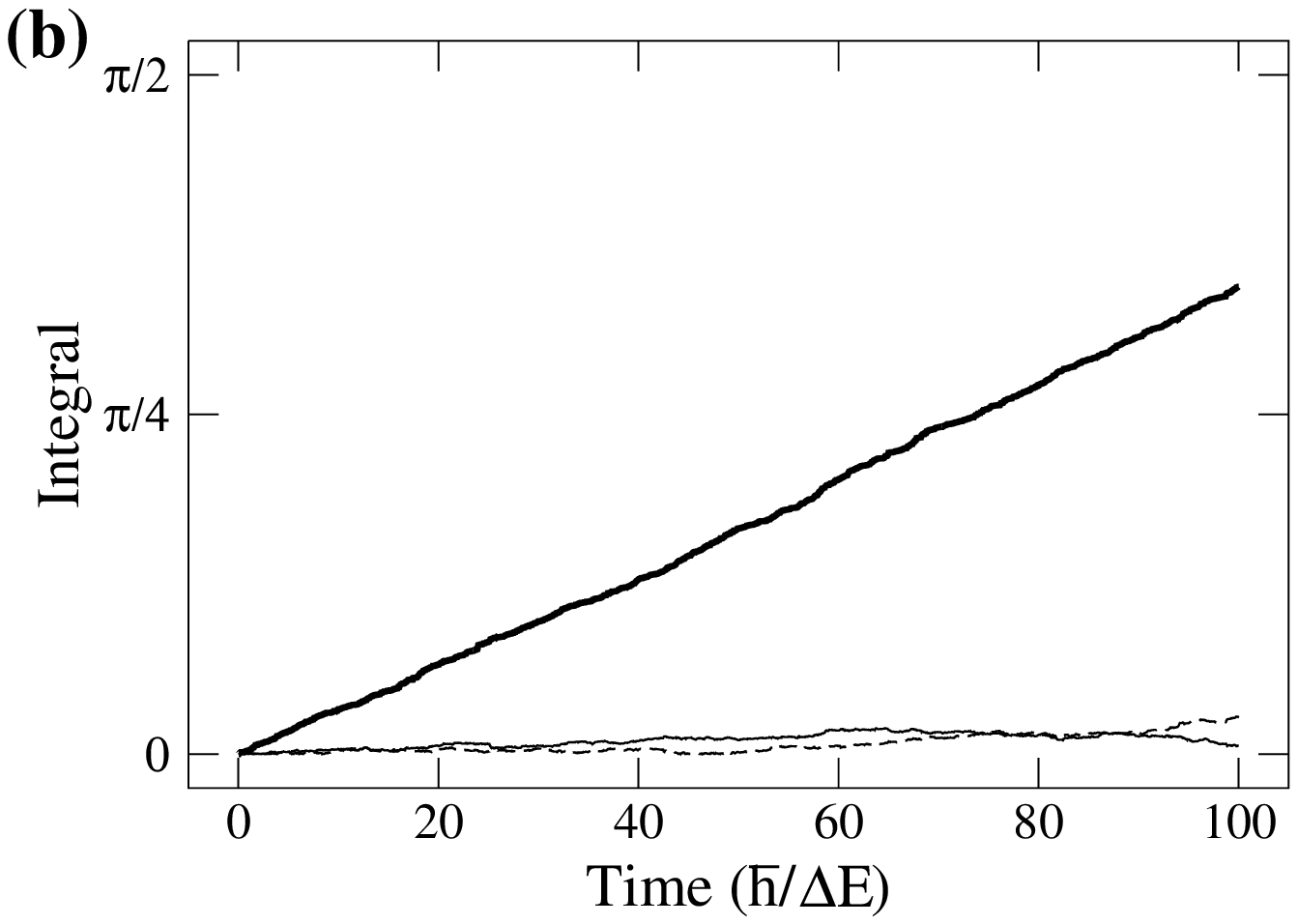}
\caption{\label{fig:oct-integral}
Time evolution of $|I_\Omega(t)|$ (thick line),
$|I_\phi(t)|$ (solid line), and $|I_\chi(t)|$ (dashed line),
 defined by Eqs.~(\ref{eqn:integrated-omega}),
(\ref{eqn:integrated-phi}), and (\ref{eqn:integrated-chi}),
respectively.
(a) and (b) correspond to Fig.~\ref{fig:numerical-short} ($T=10$ and $\alpha=1$)
and Fig.~\ref{fig:numerical-long} ($T=100$ and $\alpha=10$), respectively.
}
\end{figure*}

\subsection{Rotating Wave Approximation}

The rotating-wave approximation (RWA) 
means dropping rapidly oscillating terms in differential equations \cite{AE87}.
Since RWA is applicable to non-degenerate multi-level systems \cite{KAS02},
we will apply RWA to Eqs.~(\ref{eqn:de-A}) and (\ref{eqn:de-B})
to solve them approximately.

To justify the use of RWA, we introduce three integrals,
\begin{eqnarray}
\label{eqn:integrated-omega}
  I_\Omega(t)&=&{1\over\hbar}\int_0^t
    \varepsilon(t)\langle\tilde\phi_0(t')|V|\tilde\chi_0(t')\rangle dt',\\
\label{eqn:integrated-phi}
  I_\phi(t)&=&{1\over\hbar}\int_0^t
    \varepsilon(t)\langle\tilde\phi_0(t')|V|\tilde\phi_0(t')\rangle dt',\\
\label{eqn:integrated-chi}
  I_\chi(t)&=&{1\over\hbar}\int_0^t
    \varepsilon(t)\langle\tilde\chi_0(t')|V|\tilde\chi_0(t')\rangle dt'.
\end{eqnarray}
Note that the generalized pulse area \cite{TF04} is represented
by $2|I_\Omega(T)|$.
The numerical results
in Figs.~\ref{fig:numerical-short} and \ref{fig:numerical-long}
are used to calculate these integrals explicitly,
and their absolute values are plotted in Fig.~\ref{fig:oct-integral}.
Since 
$I_\phi(t)$ and $I_\chi(t)$ are almost zero, 
it is appropriate to assume that
\begin{equation}
\label{eqn:approx1}
  I_\phi(t)=I_\chi(t)=0,
\end{equation}
which corresponds to RWA.
Furthermore, for a long target time, we can employ the following 
form
\begin{equation}
\label{eqn:approx2}
  I_\Omega(t)= \Omega t,
\end{equation}
which means that, with use of RWA, the integrand in Eq.~(\ref{eqn:integrated-omega}) is 
nearly constant in time. 

Under these approximations (\ref{eqn:approx1}) and (\ref{eqn:approx2}),
the differential equations, (\ref{eqn:de-A}) and (\ref{eqn:de-B}),
are simplified to
\begin{equation}
  {d\over dt}A(T)=-i\Omega B(t),\quad
  {d\over dt}B(T)=-i\Omega^* A(t).
\end{equation}
For the initial value problem with the conditions, $A(0)=1$ and $B(0)=0$,
we obtain a solution,
\begin{equation}
  A(t)=\cos[|\Omega|t],\quad
  B(t)=-{i|\Omega|\over\Omega}\sin[|\Omega|t].
\end{equation}
Defining the phase of $\Omega$ as 
\begin{equation}
\label{eqn:phase-Omega}
  {\Omega\over|\Omega|}=e^{i\theta},
\end{equation}
we finally obtain
\begin{widetext}
\begin{equation}
\label{eqn:forward}
  |\phi(t)\rangle
  =|\tilde\phi_0(t)\rangle\cos[|\Omega|t]
    -ie^{-i\theta}|\tilde\chi_0(t)\rangle\sin[|\Omega|t]
  ={\cos[|\Omega|t+\Theta]\over\cos\Theta}|\phi_0(t)\rangle
    -{ie^{-i\theta}\sin[|\Omega|t]\over\cos\Theta}|\chi_0(t)\rangle.
\end{equation}
This state oscillates with the frequency $|\Omega|$
between $|\tilde\phi_0(t)\rangle$ and $|\tilde\chi_0(t)\rangle$,
as well as 
between $|\phi_0(t)\rangle$ and $|\chi_0(t)\rangle$.

\subsection{Analytic Field\label{sec:analytic-field}}

\begin{figure*}
 \includegraphics[scale=0.5]{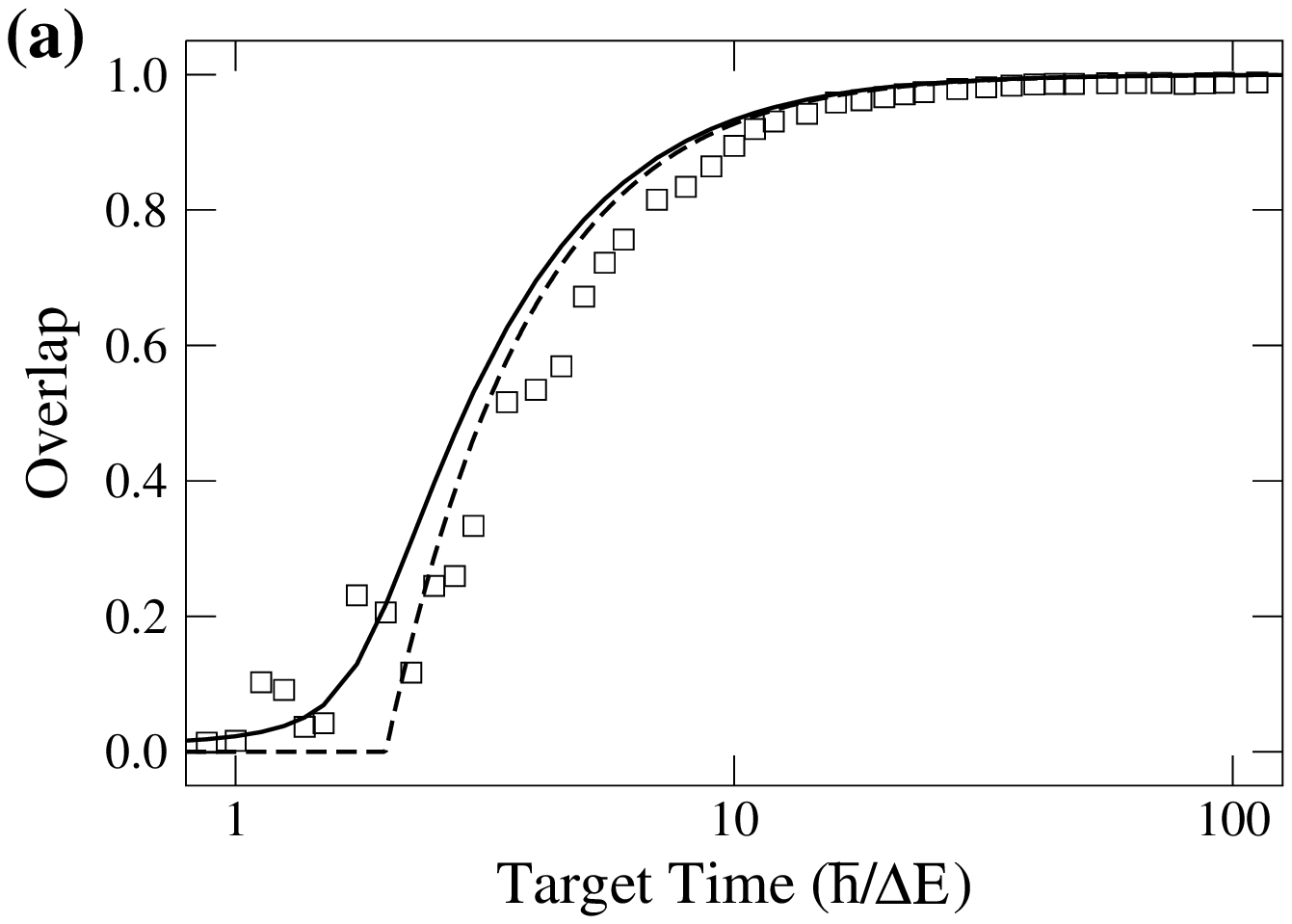}
 \includegraphics[scale=0.5]{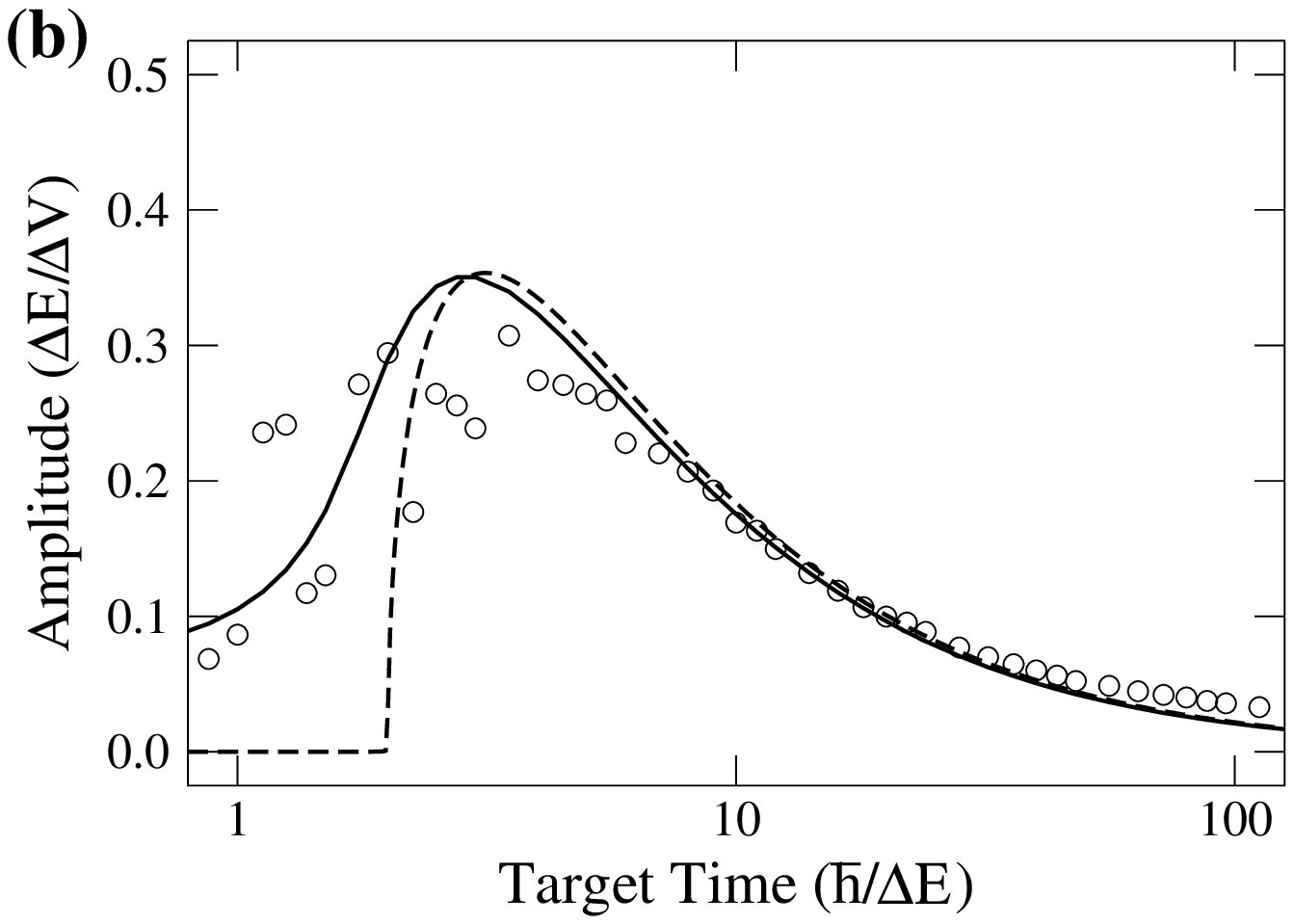}
\caption{\label{fig:oct-short}
Numerical results (squares and circles) by OCT
with the penalty factor $\alpha=1$ are
compared to our theoretical values:
(a) Final overlap $J_0$. 
(b) Averaged amplitude $\bar\varepsilon$
of the external field.
The solid (dashed) curve corresponds 
to the case of $\sin\Theta=\sqrt{\pi/4N}$ ($\Theta=0$).
}
\end{figure*}

\begin{figure*}
 \includegraphics[scale=0.5]{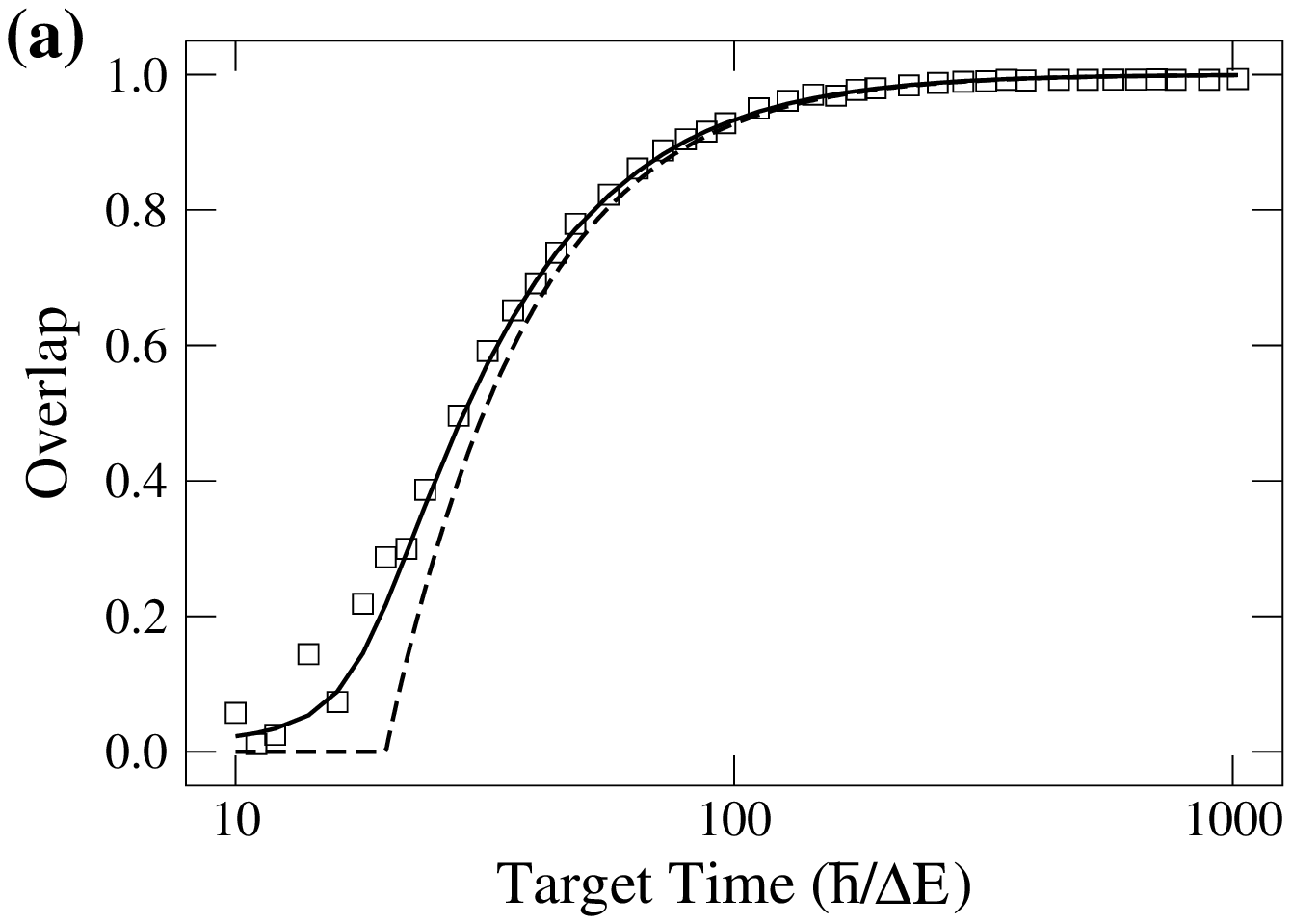}
 \includegraphics[scale=0.5]{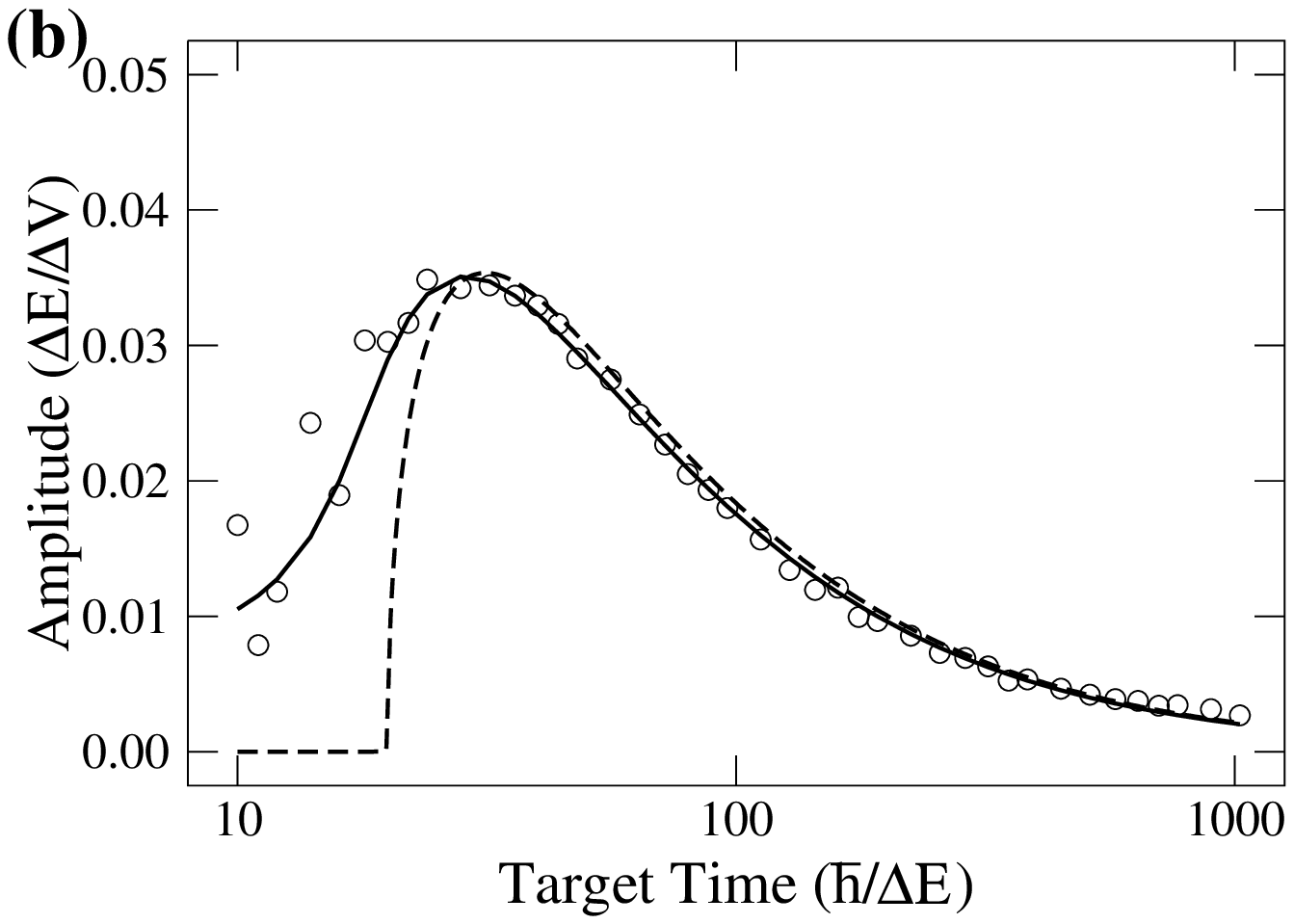}
\caption{\label{fig:oct-long}
The same as Fig.~\ref{fig:oct-short} except
that the penalty factor is larger ($\alpha=10$).
}
\end{figure*}

According to the ZBR-OCT scheme in Sec.~\ref{sec:ZBR-OCT},
the optimal external field is represented by Eq.~(\ref{eqn:ZBR-field})
given the forward evolving state $|\phi(t)\rangle$ and
the backward evolving state $|\chi(t)\rangle$ are prepared 
with the boundary conditions,
\begin{equation}
  |\phi(0)\rangle=|\Phi_0\rangle,\qquad
  |\chi(T)\rangle=|\Phi_T\rangle.
\end{equation}
Under the approximations (\ref{eqn:approx1}) and (\ref{eqn:approx2}),
$|\phi(t)\rangle$  has been already given in Eq.~(\ref{eqn:forward}),
and $|\chi(t)\rangle$ is written as
\begin{eqnarray}
\label{eqn:backward}
  |\chi(t)\rangle
  &=&-ie^{i\theta}|\tilde\phi_0(t)\rangle\sin[|\Omega|(t-T)-\Theta]
   +|\tilde\chi_0(t)\rangle\cos[|\Omega|(t-T)-\Theta]\\
  &=&-{ie^{i\theta}\sin[|\Omega|(t-T)]\over\cos\Theta}|\phi_0(t)\rangle
   +{\cos[|\Omega|(t-T)-\Theta]\over\cos\Theta}|\chi_0(t)\rangle.
\end{eqnarray}
\end{widetext}

The overlap between these states is
\begin{equation}
\label{eqn:overlap}
  \langle\phi(t)|\chi(t)\rangle=ie^{i\theta}\sin[|\Omega|T+\Theta].
\end{equation}
Substituting Eqs.~(\ref{eqn:forward}) and (\ref{eqn:backward})
into $|\phi(t)\rangle$ and $|\chi(t)\rangle$ in Eq.~(\ref{eqn:ZBR-field}),
we obtain the external field,
\begin{equation}
\label{eqn:field-final}
  \varepsilon(t)
  ={\sin[2(|\Omega|T+\Theta)]\over2\alpha\hbar}{\rm Re}\left[
    e^{i\theta}\langle\tilde\chi_0(t)|V|\tilde\phi_0(t)\rangle
  \right].
\end{equation}
By calculating $I_\Omega(T)$ for the external field (\ref{eqn:field-final})
and using Eq.~(\ref{eqn:approx2}) at $t=T$,
we obtain an equation for $\Omega$,
\begin{equation}
\label{eqn:Omega-final}
  \Omega={\sin[2(|\Omega|T+\Theta)]\over4\alpha\hbar^2}
  \left[e^{i\theta}\bar V^2+e^{-i\theta}\bar W^2\right]
\end{equation}
where $\bar V^2$ and $\bar W^2$ are defined by averages of transition elements,
\begin{eqnarray}
\label{eqn:avr-V}
  \bar V^2&=&{1\over T}\int_0^T
    \left|\langle\tilde\phi_0(t)|V|\tilde\chi(t)\rangle\right|^2dt,\\
\label{eqn:avr-W}
  \bar W^2&=&{1\over T}\int_0^T
    \left[\langle\tilde\phi_0(t)|V|\tilde\chi(t)\rangle\right]^2dt.
\end{eqnarray}
Note that $|\bar W^2|$ becomes small compared to $\bar V^2$ when
the system is sufficiently large without special symmetry
(see Eqs. (\ref{eqn:V2}) and (\ref{eqn:W2})).
Then, we obtain an equation for $|\Omega|$
with use of Eq.~(\ref{eqn:phase-Omega}),
\begin{equation}
\label{eqn:Omega-equation}
  |\Omega|={\bar V^2\sin[2(|\Omega|T+\Theta)]\over4\alpha\hbar^2}.
\end{equation}

The solutions of this equation are obtained from
the crossing points between
$y=x/T$ and $y=(K/2)\sin[2(x+\Theta)]$ where 
$x\equiv|\Omega|T$ and $K \equiv\bar V^2/(2\alpha\hbar^2)$.

To illustrate the effectiveness of our result, 
we calculate the final overlap $J_0$ from Eq.~(\ref{eqn:overlap})
\begin{equation}
\label{eqn:final-overlap}
  J_0=\left|\langle\phi(t)|\chi(t)\rangle\right|^2
     =\sin^2[|\Omega|T+\Theta]
\end{equation}
and 
the average amplitude of the optimal field from Eq.~(\ref{eqn:field-final})
\begin{equation}
  \bar\varepsilon=\sqrt{{1\over T}\int_0^T|\varepsilon(t)|^2dt}
  \simeq{\sqrt2\hbar|\Omega|\over\bar V}.
\end{equation}
These estimates match well with the numerical results (squares and circles)
as shown in Figs.~\ref{fig:oct-short} and \ref{fig:oct-long}.
Since the inner product between $N$ dimensional random complex vectors
is $\sim \sqrt{\pi/4N}$ as an average (see Eq.~(\ref{eqn:inner2})),
we have used $|\langle\phi_0(t)|\chi_0(t)\rangle|
=\sin\Theta\simeq\sqrt{\pi/4N}$ (solid lines).
For comparison, we also show the results for $\Theta=0$ (dashed lines),
which correspond to our previous results \cite{TF04,TFM05}. It is obvious that 
the new analytic field outperforms the previous one.

\begin{figure*}
 \includegraphics[scale=0.5]{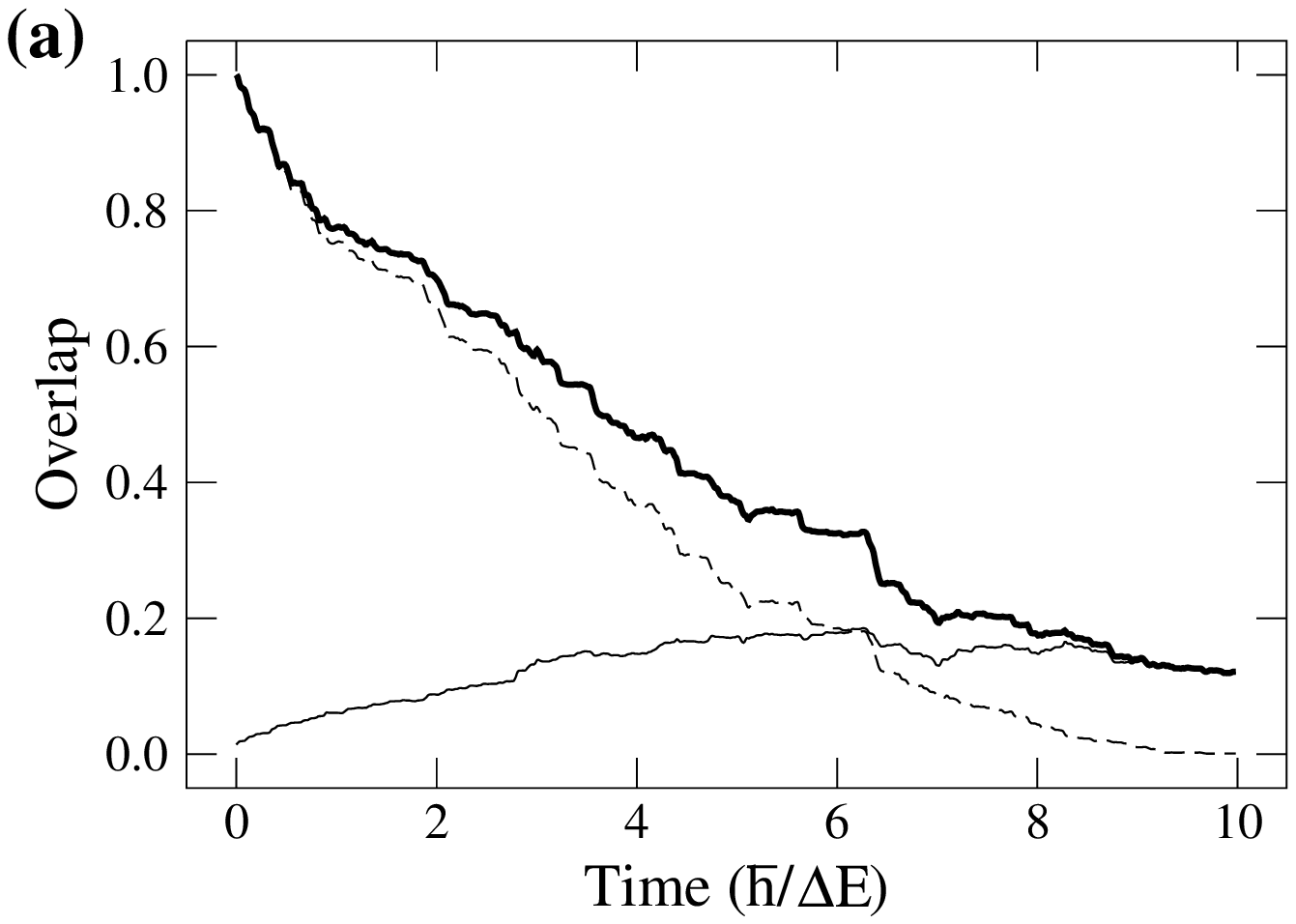}
 \includegraphics[scale=0.5]{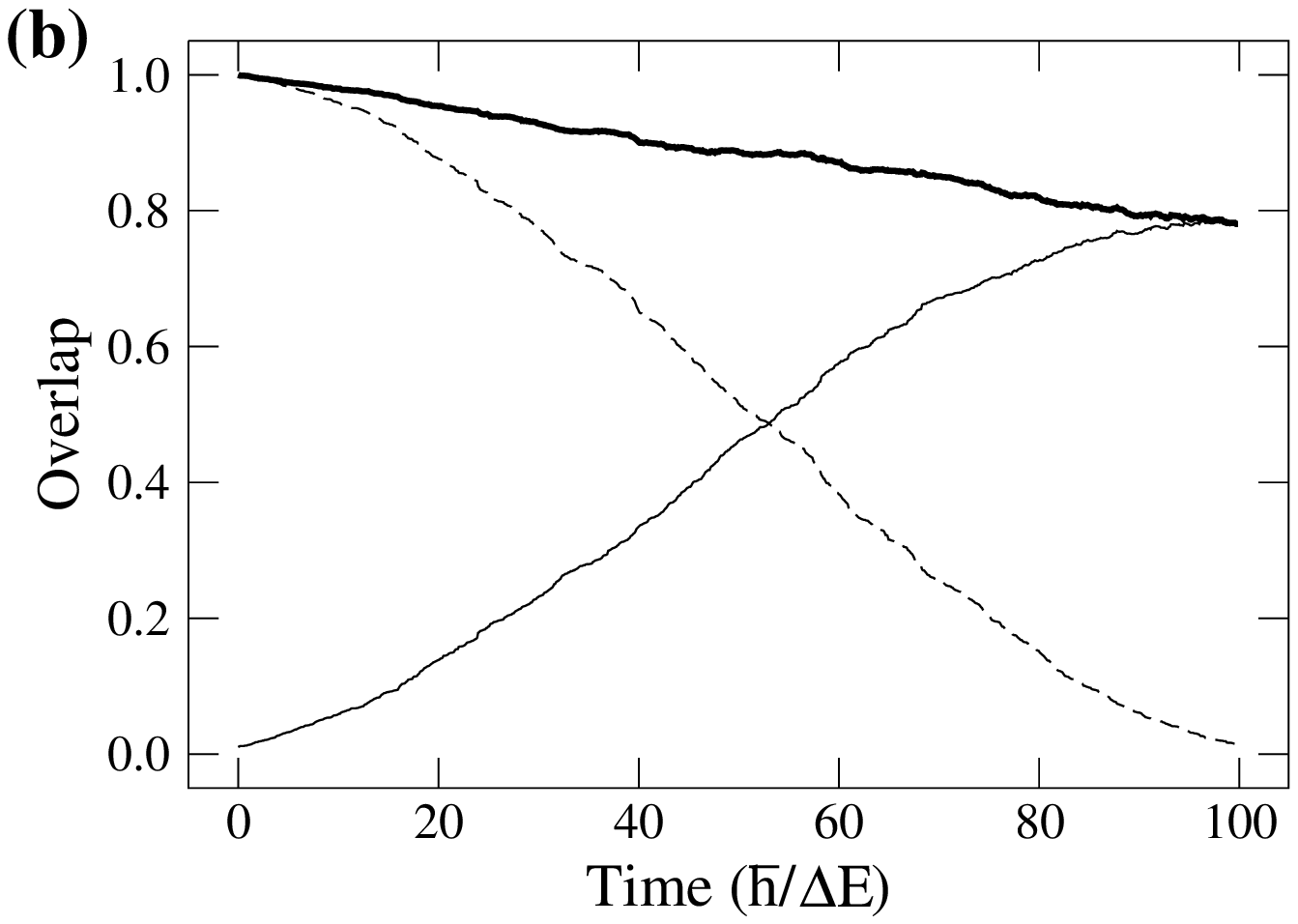}
\caption{\label{fig:analytic-smooth}
The same as Fig.~\ref{fig:numerical-smooth} except that 
we have used the analytic 
optimal field Eq.~(\ref{eqn:perfect-field}).
(a) and (b) correspond to $T=10$ and $T=100$, respectively.
}
\end{figure*}

\subsection{Perfect Control}

In the ZBR-OCT scheme (Sec.~\ref{sec:ZBR-OCT}), 
we need a finite value of the penalty factor $\alpha$
in order to avoid numerical unstability.
In the analytic approach, however, the limit $\alpha\rightarrow0$ can
be taken safely in Eq.~(\ref{eqn:Omega-equation}).
In this case, the solution of Eq.~(\ref{eqn:Omega-equation}) becomes
\begin{equation}
  |\Omega|\rightarrow\Omega_m={1\over T}\left({\pi\over2}-\Theta+m\pi\right),
\end{equation}
where $m$ is an integer.
It is easily shown that this gives $J_0=1$ from Eq.~(\ref{eqn:final-overlap}).
From Eqs.~(\ref{eqn:field-final}) and (\ref{eqn:Omega-equation}),
the external field is obtained as 
\begin{equation}
\label{eqn:perfect-field}
  \varepsilon(t)={2\hbar\Omega_m\over\bar V^2}{\rm Re}\left[
    e^{i\theta}\langle\tilde\chi_0(t)|V|\tilde\phi_0(t)\rangle
  \right].
\end{equation}
Since this equation does not contain $\alpha$,
it is different from other non-iterative optimal fields \cite{ZR99}.
It can be shown that this field actually achieves perfect control
in the limit of $T,\ N\rightarrow\infty$.
We give the proof in Appendix~\ref{sec:proof}.

\subsection{Application of the Analytic Field\label{sec:application}}

Although it is theoretically exact for $T,\ N\rightarrow\infty$,
our approach is applicable
to the cases with finite $T$ and $N$ as already 
given in Figs.~\ref{fig:oct-short} and \ref{fig:oct-long}.

\begin{figure}
\begin{center}
 \includegraphics[scale=0.5]{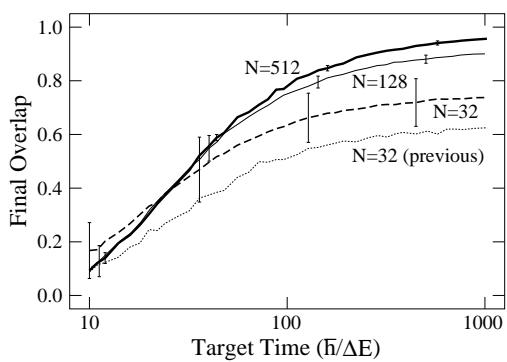}
\caption{\label{fig:performance}
The final overlap $J_0$ obtained by the analytic optimal field.
We show the results for various system sizes
according to the following procedure:
At first, we generate two random matrices ($H_0$ and $V$) and
two random vectors ($|\Phi_0\rangle$ and $|\Phi_T\rangle$)
for a system size $N$.
If we choose a target time $T$, the analytic field is given by
Eq.~(\ref{eqn:perfect-field}).
The quantum state $|\psi(t)\rangle$ at $t=T$ is obtained by
numerical integration of Schr\"odinger's equation under the field
with the initial condition $|\psi(0)\rangle=|\Phi_0\rangle$.
Finally, the final overlap $J_0$ is calculated
by $|\langle\Phi_T|\psi(T)\rangle|^2$.
Each curve in the figure is obtained as an ensemble average over
100 different realizations of random numbers.
Compare this with Fig.~1 in the previous study \cite{TF04}
(only the data of $N=32$ is shown as a dotted curve in this figure
for comparison).
}
\end{center}
\end{figure}

In Fig.~\ref{fig:analytic-smooth}, we show time evolutions of
the overlaps with $T=10$ (a) and $T=100$ (b).
The initial and target states are the same as  
in Fig.~\ref{fig:numerical-smooth}.
Unlike the results by OCT (Fig.~\ref{fig:numerical-smooth}), 
the probability on the subspace spanned by
$|\phi_0(t)\rangle$ and $|\chi_0(t)\rangle$ decreases monotonically.
This is because the analytic field (\ref{eqn:perfect-field}) was
obtained under the conditions $T,\ N\rightarrow\infty$
while the numerical calculations were performed for finite $T$ and $N$.

The performance of the analytic optimal field can be easily 
seen by plotting the final overlap $J_0$ for various values of $N$ and $T$
(Fig. \ref{fig:performance}).
The errorbars in this figure represent the normal deviation of $J_0$
obtained from calculations for 100 different samples of the Hamiltonian
and state vector.
By comparing with our previous result, Fig.1 in \cite{TF04},
our new analytic field outperforms the previous one for the 
intermediate $T$ and $N$ as expected.
This is a nice feature when we consider the application of this 
method to non-limiting cases.

\section{Summary and Discussion\label{sec:summary}}

In this paper,
we have proposed an analytic approach for controlling quantum states
in random matrix systems.
From the analysis of OCT calculations,
we showed that optimally controlled states
remain on a subspace spanned by two ``moving bases''
when the target time and system size are both sufficiently large.
According to this observation,
we developed a new method to solve OCT equations
and to obtain an analytic expression for the optimal field.
Finally, it was numerically shown that the analytic field
actually steers the quantum states in random matrix systems.
The difference from our previous result is that we have taken 
new moving bases which are exactly orthogonal, and the 
newly obtained analytic field outperforms the previous 
one for the intermediate target time and system size.

Our analytic field (\ref{eqn:perfect-field}) is
a generalized $\pi$-pulse \cite{TF04} in multi-level systems,
which is realized because of certain randomness
in the elements of the Hamiltonian and state vectors.
The amplitude of the pulse becomes smaller when the target time $T$ is larger
since an effective pulse area should be a constant $\pi$ \cite{TF04}.
Although our controlled dynamics seems to be antithetical  
to the molecular processes induced by intense laser fields 
\cite{Yamanouchi02,YS04}, 
this does not necessarily mean 
that our approach is not applicable to those systems. 
This is because such dynamics driven by the intense laser field 
can be included in the ``unperturbed'' Hamiltonian $H_0$. 
If the system becomes strongly chaotic by the laser field, 
such a situation is even preferable for the prerequisite  
of our approach using random matrix Hamiltonians. 

The quantum targeting problem studied in this paper
was solved analytically for random matrix systems,
while it is known that classical targeting problems \cite{Ott02,SR91}
are difficult to be solved for strongly chaotic cases.
This is because there is sensitivity of trajectories with respect to initial values.
Our result for quantum systems thus seems
to break the naive quantum-classical correspondence.
It is important to clarify how the correspondence is recovered in 
the semiclassical limit \cite{FTKN03,KMN05}.

Recently Gong and Brumer showed that coherent control works
for a quantized kicked rotor \cite{GB01-PRL,GB05},
a typical ``quantum chaos'' system,
whereas our concern was optimal control of quantum states in 
random matrix systems.
Optimal control for quantum chaos systems, especially weakly chaotic systems,
is another interesting subject 
which should be pursued \cite{TFM05,TF06}.

\section*{Acknowledgements}

The authors thank Prof.~S.A.~Rice, Prof.~H.~Rabitz, Prof.~M.~Toda,
Prof.~H.~Nakamura, Prof.~H.~Kono, Prof.~S.~Tasaki, 
Prof.~A.~Shudo, Dr.~Y.~Ohtsuki, and Dr.~G.V.~Mil'nikov
for useful discussions.

\appendix
\section{Eigenstate Representation\label{sec:ES-rep}}

\subsection{Preliminary}

Though our results in the main text 
do not depend on a particular representation,
in this appendix, 
we describe the controlled dynamics and the analytic optimal field
Eq.~(\ref{eqn:perfect-field})
by using the eigenstate representation of $H_0$,
and prove that perfect control is achieved by Eq.~(\ref{eqn:perfect-field}).

We introduce eigenstates $|\varphi_j\rangle$ of $H_0$ corresponding to
eigenvalues $E_j$ where
$\{|\varphi_j\rangle\}$ constitute an orthonormal basis set.
The initial and target states
\begin{equation}
  |\Phi_0\rangle
  =\sum_{j=1}^Nc_j|\varphi_j\rangle,\qquad
  |\Phi_T\rangle
  =\sum_{j=1}^Nd_j|\varphi_j\rangle
\end{equation}
are represented by random complex numbers $\{c_j\}$ and $\{d_j\}$ satisfying
normalization conditions,
\begin{equation}
\label{eqn:normalize}
  \sum_{j=1}^N|c_j|^2=\sum_{j=1}^N|d_j|^2=1.
\end{equation}
The matrix elements of $V$ are  defined by using $\{|\varphi_j\rangle\}$
\begin{equation}
  V_{jk}=\langle\varphi_j|V|\varphi_k\rangle.
\end{equation}
These quantities, $\{c_j\}$, $\{d_j\}$ and $\{V_{jk}\}$, are
assumed to be uncorrelated among them.

\subsection{The Analytic Optimal Field}

The moving bases, Eq.~(\ref{eqn:time-dependent-states}), 
satisfying $|\phi(0)\rangle=|\Phi_0\rangle$ and
$|\chi(T)\rangle=|\Phi_T\rangle$ can be written as
\begin{eqnarray}
  |\phi_0(t)\rangle
  &=&\sum_{j=1}^Nc_j|\varphi_j\rangle e^{E_jt/i\hbar},\\
  |\chi_0(t)\rangle
  &=&\sum_{j=1}^Nd_j|\varphi_j\rangle e^{E_j(t-T)/i\hbar},
\end{eqnarray}
where $T$ represents the target time.
In general, these states are not orthogonal to each other
as shown in Eq.~(\ref{eqn:inner-product}),
%
and the orthogonal (new) moving bases,
Eqs.~(\ref{eqn:tilde_phi}) and (\ref{eqn:tilde_chi}),
are constructed as
\begin{eqnarray}
  |\tilde\phi_0(t)\rangle&=&\sum_{j=1}^N
    c_j|\varphi_j\rangle e^{E_jt/i\hbar},\\
  |\tilde\chi_0(t)\rangle&=&\sum_{j=1}^N
    {d_je^{-E_jT/i\hbar}-ie^{i\theta}c_j\sin\Theta\over\cos\Theta}
      |\varphi_j\rangle e^{E_jt/i\hbar}\nonumber\\
    &\equiv&\sum_{j=1}^N\tilde d_j|\varphi_j\rangle e^{E_jt/i\hbar}.
\end{eqnarray}
Substituting these states into Eq.~(\ref{eqn:perfect-field}),
we obtain the eigenstate representation of the analytic optimal field,
\begin{equation}
\label{eqn:perfect-field-er}
  \varepsilon(t)={2\hbar\Omega_m\over\bar V^2}
    {\rm Re}\left[
      e^{i\theta}\sum_{j=1}^N\sum_{k=1}^N
       {\tilde d}_j^*V_{jk}c_k\exp\left\{{(E_k-E_j)t\over i\hbar}\right\}
  \right].
\end{equation}

\subsection{Sum of Random Variables}

Suppose a probability variable $Y$ is defined
by a sum of independent probability variables $X_j$ ($j=1,\ldots,n$),
\begin{equation}
  Y=X_1+X_2+\cdots+X_n,
\end{equation}
with an expectation ${\mathcal M}(X_j)={\mathcal M}_X$ ($>0$)
and a variance $\sigma^2(X_j)={\sigma_X}^2$.
When $n$ is sufficiently large,
it is known from the central limit theorem
that $Y$ is normally distributed. 
The expectation and variance are 
\begin{equation}
  {\mathcal M}(Y)=n{\mathcal M}_X,\qquad\sigma^2(Y)=n{\sigma_X}^2.
\end{equation}
Then, the expectation $n{\mathcal M}_X$ can be used as an approximate value of $Y$
since the relative standard deviation
\begin{equation}
  \sigma_{\rm rel}
    \equiv\frac{\sqrt{\sigma^2(Y)}}{{\mathcal M}(Y)}
    =\frac{\sigma_X}{{\mathcal M}_X\sqrt n}\simeq O(1/\sqrt n)
\end{equation}
vanishes for $n\rightarrow\infty$.

Using the above basic knowledge, we estimate approximated values
for sums of $V_{kj}$, $c_j$, and $\tilde d_k$ in the following.
The coefficients $c_j$ and $\tilde d_j$ are independent random numbers
subject to the distribution function (\ref{eqn:random-vector}), i.e.,
\begin{eqnarray}
  {\mathcal M}(|c_j|^2)&=&{\mathcal M}(|\tilde d_j|^2)=\frac1N,\\
  \sigma^2(|c_j|^2)&=&\sigma^2(|\tilde d_j|^2)=\frac1{N^2}.
\end{eqnarray}

For large $N$, we obtain
\begin{eqnarray}
\label{eqn:reduction1}
  \sum_{j=1}^N|V_{kj}|^2|c_j|^2
    &\simeq&N\ {\mathcal M}(|V_{kj}|^2|c_j|^2)\nonumber\\
    &=&{\mathcal M}(|V_{kj}|^2)+O(1/\sqrt N),\\
\label{eqn:reduction2}
  \sum_{k=1}^N|\tilde d_k|^2|V_{kj}|^2
    &\simeq&N\ {\mathcal M}(|\tilde d_k|^2|V_{kj}|^2)\nonumber\\
    &=&{\mathcal M}(|V_{kj}|^2)+O(1/\sqrt N),
\end{eqnarray}
where we have used a basic relation
\begin{equation}
  {\mathcal M}(X_1X_2\cdots X_n)
  ={\mathcal M}(X_1){\mathcal M}(X_2)\cdots {\mathcal M}(X_n)
\end{equation}
for independent probability variables $X_j$.

Applying the central limit theorem
to a sum of complex variables $Z_n=z_1+z_2+\cdots+z_n\equiv X_n+iY_n$ with
${\mathcal M}(z_j)=0$ and ${\mathcal M}(|z_j|^2)={\sigma_z}^2$,
we have
\begin{eqnarray}
  &&P(X_n)P(Y_n)\ dX_ndY_n\nonumber\\
  &&=\frac1{\pi n{\sigma_z}^2}
     \exp\left(-\frac{{X_n}^2+{Y_n}^2}{n{\sigma_z}^2}\right)
     \ dX_ndY_n.
\end{eqnarray}
Then, the average magnitude of $|Z_n|$ can be calculated as
\begin{equation}
\label{eqn:central-limit}
  {\mathcal M}(|Z_n|)
   =\int|Z_n|P(X_n)P(Y_n)\ dX_ndY_n=\frac{\sqrt{\pi n{\sigma_z}^2}}{2}.
\end{equation}
From this relation,
the inner product, Eq.~(\ref{eqn:inner-product}), is estimated as
\begin{equation}
  \left|\langle\phi_0(t)|\chi_0(t)\rangle\right|
  \simeq\frac{\sqrt{\pi N\ {\mathcal M}(|c_j|^2|d_j|^2)}}{2}
  =\sqrt\frac{\pi}{4N}.
  \label{eqn:inner2}
\end{equation}

In the same manner, the average transition elements
(\ref{eqn:avr-V}) and (\ref{eqn:avr-W}) are estimated as
\begin{widetext}
\begin{eqnarray}
\label{eqn:V2}
  \bar V^2
   &=&\left|\sum_jc_j^*V_{jj}\tilde d_j\right|^2
     +\sum_j\sum_{k\ne j}\left|c_j^*V_{jk}\tilde d_k\right|^2
   \simeq{\mathcal M}(|V_{kj}|^2)+O(1/N),\\
\label{eqn:W2}
  \bar W^2
   &=&\left(\sum_jc_j^*V_{jj}\tilde d_j\right)^2
     +2\sum_j\sum_{k<j}c_j^*c_k^*\left|V_{jk}\right|^2\tilde d_j\tilde d_k
   \simeq O(1/N),
\end{eqnarray}
\end{widetext}
in the limit $T\rightarrow\infty$.
Thus, we can ignore $\bar W^2$ for $T,\ N\rightarrow\infty$.

\subsection{Controlled State\label{sec:proof}}

We give a proof that a quantum state driven by
the analytic field, Eq.~(\ref{eqn:perfect-field}) or (\ref{eqn:perfect-field-er}), 
actually shows a smooth transition between the initial and target states.
We assume that the size $N$ of the random matrix Hamiltonian $H_0$ and
the target time $T$ are both large enough.

To see the dynamics induced by the field,
we represent a quantum state in the eigenstate representation,
\begin{equation}
  |\psi(t)\rangle=\sum_ja_j(t)|\varphi_j\rangle e^{E_jt/i\hbar},
\end{equation}
which satisfies the initial condition $|\psi(0)\rangle=|\Phi_0\rangle$.
From the Schr\"odinger's equation driven by the 
optimal field (\ref{eqn:perfect-field-er}), 
we obtain the following differential equations for $a_j(t)$,
\begin{equation}
\label{eqn:deq-er}
  \dot a_k(t)=-i\sum_{j\ne k}\Omega_{kj}a_j(t),
 \end{equation}
where $\Omega_{kj}$ is defined by
\begin{equation}
  \Omega_{kj}={\Omega_m|V_{kj}|^2\over\bar V^2}\left(
      c_j^*\tilde d_k+\tilde d_j^*c_k
  \right),
\end{equation}
and we have used the rotating-wave approximation.
We write $a_k(t)$ in the following form,
\begin{equation}
  a_k(t)=A_k\cos\Omega_mt+B_k\sin\Omega_mt,
\label{eqn:deq-er2}
\end{equation}
and $A_k$ and $B_k$ are determined as
\begin{equation}
  A_k=c_k,\qquad
  B_k=-i\sum_{j\ne k}{\Omega_{kj}\over\Omega_m}c_j,
\end{equation}
from the initial conditions $a_k(0)=c_k$ and (\ref{eqn:deq-er}).

Using the relations (\ref{eqn:reduction1}) and (\ref{eqn:reduction2}),
we obtain
\begin{widetext}
\begin{eqnarray}
\label{eqn:sum1}
  \sum_{j\ne k}{\Omega_{kj}\over\Omega_m}c_j
  &=&\tilde d_k\sum_{j\ne k}\frac{|V_{kj}|^2|c_j|^2}{\bar V^2}
  +c_k\sum_{j\ne k}\frac{|V_{kj}|^2c_j\tilde d_j^*}{\bar V^2}
  \simeq\tilde d_k+O(1/N),\\
\label{eqn:sum2}
  \sum_{j\ne k}{\Omega_{kj}\over\Omega_m}\tilde d_j
  &=&\tilde d_k\sum_{j\ne k}\frac{|V_{kj}|^2c_j^*\tilde d_j}{\bar V^2}
  +c_k\sum_{j\ne k}\frac{|V_{kj}|^2|\tilde d_k|^2}{\bar V^2}
  \simeq c_k+O(1/N).
\end{eqnarray}
\end{widetext}
Substituting (\ref{eqn:sum1}) into (\ref{eqn:deq-er2}), 
we obtain
\begin{equation}
\label{eqn:a_k}
  a_k(t)=c_k\cos\Omega_mt-i\tilde d_k\sin\Omega_mt,
\end{equation}
and the right-hand side of the differential equation (\ref{eqn:deq-er})
becomes
\begin{equation}
\label{eqn:rhs}
  -i\sum_{j\ne k}\Omega_{kj}a_j(t)=-\Omega_m\left(
    c_k\sin\Omega_mt+i\tilde d_k\cos\Omega_mt
  \right)
\end{equation}
with use of  (\ref{eqn:sum1}) and (\ref{eqn:sum2}).
Since Eq.~(\ref{eqn:rhs}) is exactly the same as $\dot a_k(t)$,
we have confirmed that (\ref{eqn:a_k}) 
is the solution for the Schr\"odinger's equation 
driven by the analytical optimal field.

The final expression (\ref{eqn:a_k}) shows that
each $a_k(t)$ smoothly changes its value from $c_k$
at $t=0$ to $\tilde d_k$ at $t=T$ as expected, and
the overlap between $|\psi(t) \rangle$ and 
$|\chi_0(t)\rangle$ is easily calculated as
\begin{equation}
  \langle\chi_0(t)|\psi(t)\rangle=-i\sin\Omega_mt.
\end{equation}
This shows $|\langle\Phi_T|\psi(T)\rangle|=1$, i.e.
perfect control is accomplished at the target time $T$.

\bibliography{BibTeX/General,BibTeX/QuantumChaos,BibTeX/Control,BibTeX/MyWorks}

\end{document}